\documentclass[prd,aps,tightenlines,superscriptaddress,nofootinbib,preprint]{revtex4}
\usepackage{amsthm}

\theoremstyle{remark}

\usepackage{graphicx}
\usepackage{dcolumn}
\usepackage{bm}
\usepackage{color}
\usepackage{amsmath}
\usepackage{amsfonts}
\usepackage{comment}
\usepackage{amssymb}
\usepackage{orcidlink}
\hypersetup{
    colorlinks=true,
    linkcolor=blue,
    linktoc=page
}\begin{document}


\title{On the Maldacena–Shenker–Stanford chaos bound in black hole spacetimes: Saturation, apparent violations, and their geometric origin}

\author{Terkaa Victor Targema \orcidlink{0000-0001-8809-1741} \footnote{{\color{purple}s2571502@ipc.fukushima-u.ac.jp,~terkaa.targema@tsuniversity.edu.ng}}}
\affiliation{Faculty of Symbiotic Systems Science, Fukushima University, Fukushima 960-1296, Japan}
\affiliation{Department of Physics, Taraba State University, Jalingo, Nigeria}

\author{Kazuharu Bamba \orcidlink{0000-0001-9720-8817}
\footnote{{\color{purple}bamba@sss.fukushima-u.ac.jp}}}
\affiliation{Faculty of Symbiotic Systems Science, Fukushima University, Fukushima 960-1296, Japan}
\author{Usman Zafar \orcidlink{0000-0001-9610-1081} \footnote{{\color{purple}s2471001@ipc.fukushima-u.ac.jp,~zafarusman494@gmail.com}}}
\affiliation{Faculty of Symbiotic Systems Science, Fukushima University, Fukushima 960-1296, Japan}

\begin{abstract}
Reported violations of the Maldacena-Shenker-Stanford (MSS) chaos bound in black hole spacetimes raise the question of whether they reflect a genuine breakdown, a feature of particular geometries, or a limitation of the probe used to test the bound. We show that the violations are apparent and arise predominantly in the near-extremal regime, where the black hole develops an $AdS_2\times S^2$ throat, the surface gravity becomes parametrically small, while the Lyapunov exponent of an orbit separated from the throat remains nonzero. We establish a geometric criterion governing this behavior and demonstrate its resolution in two complementary ways: a charged-particle construction that localizes an otherwise excluded orbit within the throat and restores exact saturation, and an out-of-time-order correlator (OTOC) calculation in the Jackiw-Teitelboim (JT) throat, which identifies the near-horizon Lyapunov scale and shows why the photon-sphere exponent does not reproduce it when the orbit remains outside the throat. We illustrate the criterion for static and rotating black holes in Einstein, scalar-tensor, and higher-curvature gravity, representative of a broader class of geometries to which it applies. In the absence of external effects that displace the relevant trajectories from the throat, the MSS bound remains respected; the apparent violations instead reflect a geometric decoupling between the probe and the near-horizon region.
\end{abstract}
                             
\maketitle


\section{Introduction}
\label{introduction}
The strong–field regime of gravity provides a natural arena for testing extensions of general relativity (GR). Black hole spacetimes furnish geometrically clean backgrounds in which departures from Einstein gravity manifest directly in the structure of geodesic motion. In theories involving additional fields or nonminimal curvature couplings, the spacetime geometry departs from the Einstein solution, thereby reshaping the orbital structure as test particles follow geodesics of the resulting metric. The existence and stability of circular orbits are determined by local curvature properties of the spacetime. 

In classical dynamical systems, such as black hole geodesic motion, chaos is typically associated with exponential sensitivity to infinitesimal perturbations of initial conditions, leading to the divergence of nearby trajectories in phase space. Unstable circular geodesics provide a natural setting in which this behavior emerges: small radial perturbations grow exponentially with time, reflecting an intrinsic dynamical instability. This behavior is quantified by the Lyapunov exponent, $\lambda$, which sets the characteristic timescale governing near-horizon motion \cite{lei2022circular,gao2022chaos,jeong2023homoclinic, c2}.

In semiclassical settings, such exponential sensitivity acquires a broader significance. The instability exponent governing near–horizon trajectories has been well-known to control the early-time growth of out-of-time-ordered correlators (OTOCs), $C(t)\sim e^{2\lambda t}$, thereby establishing a correspondence between classical orbital instability and quantum chaotic dynamics \cite{deich2024lyapunov, pourkhodabakhshi2025saturation}. Within holographic frameworks, this quantum Lyapunov exponent admits a gravitational interpretation, where black hole horizons encode information-theoretic properties of the dual field theory \cite{ads1,ads2,ads3,ads4,hashimoto2022bound,prihadi2023chaos}.

A fundamental constraint on chaos in thermal quantum systems, extended to Einstein gravity via the anti-de Sitter/conformal field theory (AdS/CFT) correspondence, was established by Maldacena, Shenker, and Stanford (MSS) \cite{maldacena2016bound}, who showed that the Lyapunov exponent satisfies $\lambda \leq 2\pi T$ (in units where $\hbar=c=k_B=1$). Using the Hawking relation between the black hole temperature \(T\) and the surface gravity \(\kappa\), the MSS bound translates into the geometric inequality \(\lambda \leq \kappa\). Physically, this relation reflects the deep connection between the thermodynamic and dynamical sectors of black hole physics. By relating the Lyapunov exponent to the temperature of the system, the MSS bound links orbital instability to the thermal properties of the black hole background. In black hole geometries, these quantities arise from distinct geometric aspects: the surface gravity reflects the thermodynamics of the horizon, while the Lyapunov exponent is determined by the curvature of the effective potential governing particle orbits. The question of whether these two scales remain universally connected, particularly in extreme conditions like near-extremal black holes, is of fundamental significance. Notably, a wide class of black hole solutions within GR satisfies or saturates this bound in appropriate regimes \cite{pourkhodabakhshi2025saturation}, revealing a nontrivial interplay between horizon thermodynamics and orbital instability. However, the generality of this identification remains an open question, as departures from the MSS bound have been reported in theories with higher-spin interactions \cite{hashimoto2017universality}, additional charges, or nonminimal couplings \cite{addazi2021chaotic,xie2023circular,d2,d3,lee2025bound,d6,d9,kan2022bound,gwak2022violation,lei2024thermodynamic,fayyaz2025chaotic,gallo2025bounds}. A unifying geometric mechanism underlying these violations remains an open question, leaving it unclear whether they signal a breakdown of the MSS bound in gravitational settings or instead reflect limitations of the classical probes employed.
The issue is particularly acute near extremality, where the surface gravity vanishes while the Lyapunov exponent can remain nonzero \cite{dalui2026chaotic}. This motivates a systematic investigation into whether such violations arise from the spacetime geometry itself or from the probes used to test the bound. To this end, we consider black hole solutions across a broad range of gravitational theories—shift-symmetric Horndeski gravity, Einstein-Gauss-Bonnet (EGB) gravity, and general relativistic cases including Reissner-Nordstr\"om, Bardeen, and Kerr—allowing us to probe singular, regular, and rotating geometries within a unified framework.

Our goal is to identify the geometric conditions underlying violations of the MSS bound in black hole spacetimes, assess their implications for holographic principles in black hole physics, and determine whether they reflect genuine breakdowns or limitations in extending quantum-motivated chaos bounds to classical orbital dynamics.
The remainder of this paper is organized as follows. In Sec.~\ref{sec:Lyapunov}, we review the derivation of Lyapunov exponents for timelike and null circular orbits in static, spherically symmetric spacetimes. In Sec.~\ref{sec2}, we analyze potential apparent violations of the MSS bound in shift-symmetric Horndeski gravity, while Sec.~\ref{sec3} is devoted to the corresponding analysis in EGB gravity. In Sec.~\ref{sec4}, we identify the universal geometric mechanism underlying the reported violations, trace its origin to the general relativistic limit, and show how it reconciles the apparent conflict with the MSS bound. Finally, Sec.~\ref{sec5} summarizes our results and discusses their implications.
\section{From quantum chaos to geodesic instability}
We begin by reviewing the aspects of the MSS conjecture relevant to the present work, following Ref.~\cite{maldacena2016bound}. The MSS bound establishes a universal upper bound on information scrambling in thermal quantum systems, quantified by the OTOC,
\begin{equation}
C(t)=-\left\langle [W(t),V(0)]^2\right\rangle_\beta,
\label{otoc}
\end{equation}
where $\langle \cdot \rangle_\beta = Z^{-1}\mathrm{Tr}(e^{-\beta H} \cdot)$ denotes the thermal expectation value at inverse temperature $\beta = 1/T$ \cite{roberts2016lieb, shenker2014black, shenker2015stringy, roberts2015localized, feng2017butterfly, blake2016universal}. For a chaotic system, the OTOC exhibits early-time exponential growth, $C(t)\sim e^{2\lambda t}$, where $\lambda$ is the quantum Lyapunov exponent. In the semiclassical limit, the commutator is replaced by the Poisson bracket, $[W(t),V(0)] \simeq i\{W(t),V(0)\}$, so the OTOC becomes proportional to the classical Poisson bracket squared. 
Under the assumptions of analyticity, causality, and factorization, the MSS conjecture yields $\lambda \leq 2\pi T$. In GR, this bound is saturated through the universal near-horizon Rindler geometry \cite{maldacena2016bound}. 

Alongside the OTOC description of quantum chaos, black hole spacetimes admit a classical characterization of dynamical instability through the Lyapunov exponent of unstable circular geodesics. A standard formulation of this exponent for black hole geodesics was developed by Cardoso \textit{et al.} \cite{cardoso2009geodesic} in the context of geodesic stability and the eikonal quasinormal-mode correspondence, and has since become a standard measure of orbital instability in black hole geometries. Motivated in part by holography, whereby suitable bulk dynamics can encode the OTOC growth of a dual quantum theory, numerous works have investigated whether this classical instability exponent can be related to the MSS Lyapunov exponent. Importantly, the Lyapunov exponent of an unstable geodesic characterizes local exponential sensitivity of a classical orbit and is not, by itself, a measure of quantum many-body chaos in the sense of nonintegrability or phase-space mixing.
 The comparison with the MSS bound is instead motivated by the fact that the corresponding classical OTOC, formulated through Poisson brackets of phase-space variables, can exhibit the same exponential growth associated with the instability of the orbit. Thus, when a suitable semiclassical correspondence between the bulk classical dynamics and quantum OTOC growth is established, the geodesic Lyapunov exponent provides a natural classical probe of the MSS bound. The Cardoso-type Lyapunov exponent, which characterizes the instability of unstable circular orbits through the curvature of the effective potential, has therefore been widely employed in this context \cite{addazi2021chaotic,xie2023circular,d2,d3,lee2025bound,d6,d9,kan2022bound,gwak2022violation,lei2024thermodynamic,fayyaz2025chaotic,gallo2025bounds}. To formulate the classical analogue of the OTOC, we adopt the standard semiclassical prescription and choose the phase-space variables $W(t)=r(t)$ and $V(0)=p_r(0)$, so that the classical OTOC, $C_{\rm cl}(t)$ can be written as
\begin{equation}
C_{\rm cl}(t)\propto \{r(t),p_r(0)\}^2.
\end{equation}
Using the canonical relation $\{r(0),p_r(0)\}=1$, one obtains
\begin{equation}
\{r(t),p_r(0)\}=\frac{\partial r(t)}{\partial r(0)},
\end{equation}
so that a small perturbation satisfies
\begin{equation}
\delta r(t)=
\frac{\partial r(t)}{\partial r(0)}\,\delta r(0).
\end{equation}
For an unstable circular orbit, the radial perturbation obeys
\begin{equation}
\delta\ddot r+\frac{{\rm V}_{\rm eff}''(r_0)}{2}\delta r=0,
\end{equation}
with solution $\delta r(t)\sim e^{\lambda t}$, where
$\lambda^2=-{\rm V}_{\rm eff}''(r_0)/2$. Consequently,
$C_{\rm cl}(t)\propto e^{2\lambda t}$,
so that the geodesic Lyapunov exponent governs the exponential growth of the classical OTOC, in direct analogy with the quantum Lyapunov exponent. This correspondence provides the basis for comparing geodesic instability with the MSS chaos bound and has motivated the use of unstable black hole orbits as classical probes of chaos. Whether such a probe reproduces the near-horizon Lyapunov scale, however, depends on whether the corresponding orbit genuinely accesses the near-horizon geometry. As we shall show, this  distinction becomes particularly important in the near-extremal regime, where the emergence of a long throat can separate the orbital dynamics from the horizon scale. We therefore revisit this comparison and identify the geometric origin of the apparent violations reported in the literature.
    \subsection{Lyapunov exponent of charged particles}
\label{sec:Lyapunov}
The Lyapunov exponent provides a quantitative measure of the instability of classical trajectories and plays an important role in characterizing sensitive dependence on initial conditions.   The general framework for characterizing dynamical instability through Lyapunov exponents has been extensively developed in the literature and applied to geodesic motion in curved spacetimes (see, e.g., Refs.~\cite{cardoso2009geodesic,gallo2025bounds,gao2022chaos,gwak2022violation,fayyaz2025chaotic,xie2023circular,Targema2025MSS}). For completeness, we briefly summarize the essential ingredients underlying this formalism.

Consider a dynamical system described by the phase-space variables $X_i(t)$ satisfying
\begin{equation}
\frac{dX_i}{dt}=F_i(X),
\end{equation}
where $F_i$ specifies the equations of motion. Let $X_i(t)$ denote a reference trajectory and introduce a small perturbation $\delta X_i(t)$. The equations of motion can be linearized about the reference trajectory in the following way: 
\begin{equation}
\frac{d}{dt}\delta X_i
=
K_{ij}(t)\,\delta X_j,
\qquad
K_{ij}(t)
=
\left.
\frac{\partial F_i}{\partial X_j}
\right|_{X(t)},
\end{equation}
where $K_{ij}$ is the Jacobian (stability) matrix evaluated along the trajectory. The evolution of the perturbation can be expressed as
\begin{equation}
\delta X_i(t)
=
L_{ij}(t)\,\delta X_j(0),
\end{equation}
where the fundamental matrix $L_{ij}(t)$ satisfies
\begin{equation}
\dot{L}_{ij}
=
K_{ik}(t)L_{kj},
\qquad
L_{ij}(0)=\delta_{ij}.
\end{equation}
The largest Lyapunov exponent is obtained from the asymptotic growth rate of $L_{ij}$,
\begin{equation}
\lambda
=
\lim_{t\rightarrow\infty}
\frac{1}{t}
\ln\|L(t)\|,
\label{eqn:lambda_matrix}
\end{equation}
which measures the exponential divergence of nearby trajectories in phase space.

In the present work, we do not repeat the complete derivation of this formalism. Instead, we apply it to charged-particle motion in static, spherically symmetric spacetimes described by
\begin{equation}
    \label{hb}
    ds^2=-h(r)\,dt^2+\frac{dr^2}{f(r)}+r^2d\Omega^2\,.
\end{equation}
The motion of a charged test particle with charge $q$ in this background is governed by the Lagrangian
\begin{equation}
\label{lang}
2\mathcal{L} = -h\dot{t}^2 + \frac{1}{f}\dot{r}^2 + D\dot{\phi}^2 - 2q\Phi_{\rm t}\,\dot{t}\,,
\end{equation}
where $\Phi_{\rm t}$ is the gauge field potential of the spacetime. The corresponding Hamiltonian can be written as 
\begin{equation}
\label{h1}
\mathcal{H}
=\frac{-\bigl(p_{t}+q\Phi\bigr)^{2}
      +fh\,p_{r}^{2}
      +hD^{-1}\,p_{\phi}^{2}}
{2h}\,,
\end{equation}
with conserved energy \(E=-p_{t}\) and angular momentum \(L=p_{\phi}\).
Restricting to equatorial motion, the radial dynamics can be expressed as a first-order dynamical system
\begin{equation}
\frac{dr}{dt}=F_{1}(r,p_{r})\,,\qquad
\frac{dp_{r}}{dt}=F_{2}(r,p_{r})\,, \label{fi}
\end{equation}
whose explicit form follows from the Hamiltonian equations and the normalization of the four-velocity for particles: \( g_{\mu\nu} \dot{x}^\mu \dot{x}^\nu = \eta \), where $\eta=0,-1$ for massless and massive particles, respectively \cite{gao2022chaos, gwak2022violation}. By considering massive particles and this normalization, the dynamical Eqs. \eqref{fi} can be expressed as 
\begin{eqnarray}
\label{dp}
F_1&=&\frac{p_rfh}{\sqrt{h\left(1+p_r^2f+\frac{p_\phi^2}{D}\right)}}\,,\nonumber\\
F_2&=&-q\Phi_{\rm t}^\prime-\frac{p_r^2(fh)^\prime+h^\prime+p_\phi^2\left(\frac{h}{D}\right)^\prime}{2\sqrt{h\left(1+p_r^2f+\frac{p_\phi^2}{D}\right)}}\,.
\end{eqnarray}

For charged timelike particles, the analysis of circular orbits is fundamental to understanding the stability properties of black hole spacetimes, since such trajectories represent equilibrium configurations in which gravitational attraction, centrifugal effects, and electromagnetic interactions effectively balance.  Circular orbits located at \(r=r_{0}\) are determined by the following equilibrium conditions:
\begin{equation}
p_{r}(r_{0})=0\,,
\qquad
\left.\frac{dp_{r}}{dt}\right|_{r_{0}}=0\,, \label{pdot}
\end{equation}
which ensures that the radial velocity and acceleration vanish. Linearizing the radial dynamics around this fixed point leads to a \(2\times2\) Jacobi matrix given by 
\begin{equation*}
    \mathbf{J}
=
\begin{pmatrix}
0 & \partial_{p_r}F_1 \\
\partial_rF_2 & 0
\end{pmatrix}\,,
\end{equation*}
whose eigenvalues determine the local stability of the orbit. The Lyapunov exponent \(\lambda\) is obtained by the positive eigenvalue of this matrix that quantifies the rate at which nearby trajectories diverge, which can be expressed by the following equation:
\begin{equation}
\label{eqn: Glyapunov}
\lambda^2 =
\frac{1}{4}
\frac{f\left[ h'(r) + p_\phi^2 \left(D^{-1} h\right)' \right]^2}
{h\left(1 + p_\phi^2 D^{-1}\right)^2}-
\frac{1}{2} f
\frac{ h''(r) + p_\phi^2 \left(D^{-1} h\right)'' }
{ 1 + p_\phi^2 D^{-1} }
-
\frac{ q\, \Phi_{\rm t}''\, f\, h }
{ \sqrt{ h\left(1 + p_\phi^2 D^{-1} \right) } } \,,
\end{equation}
where all quantities are evaluated at the circular-orbit radius \(r_{0}\).
A positive real value of $\lambda^{2}$ indicates an unstable circular orbit, whereas $\lambda^{2}<0$ corresponds to stable motion. 

The angular momentum plays a crucial role in determining the orbital dynamics of particles around black holes. Accordingly, it is consistently incorporated into our analysis. A number of recent studies have shown that departures from the MSS bound are closely tied to angular momentum effects \cite{lei2024thermodynamic, d7, fayyaz2025chaotic, xie2023circular}. While obtaining closed-form expressions for the angular momentum of charged particles is generally difficult—leading many analyses to rely primarily on numerical methods—we demonstrated in Ref.~\cite{Targema2025MSS} that the angular momentum admits an exact analytical expression. The explicit form of this quantity plays a crucial role in the stability analysis, since the structure and existence of circular orbits depend sensitively on the associated angular-momentum constraints. In the present study, we employ these constraints, which are given by
\begin{equation}\label{eq:L2-general}
L_0^2
=
\frac{-\mathcal{A}_2 \pm \sqrt{\mathcal{A}_2^2 - 4 \mathcal{A}_1 \mathcal{A}_3}}
{2 \mathcal{A}_1}\,,
\end{equation}
where the coefficients are given by 
\begin{eqnarray}
\mathcal{A}_1 &=&
\left[\left(\frac{h}{D}\right)'\right]^2\,,
\label{coeff:alpha} \\[1ex]
\mathcal{A}_2 &=&
2\left(\frac{h}{D}\right)' h'
-\frac{4 q^2 \Phi_{\rm t}'\,^2 h}{D}\,,
\label{coeff:beta} \\[1ex]
\mathcal{A}_3 &=&
h'^2 - 4 q^2 \Phi_{\rm t}'\,^2 h\,.
\label{coeff:gamma}
\end{eqnarray}

Timelike circular orbits are determined from the conditions that the conserved angular momentum be real and finite. This requirement reduces to the inequality
\begin{equation}
\label{ineq}
\mathcal{A}_1(r) > 0\,,
\end{equation}
which ensures the existence of admissible timelike circular solutions. By imposing this condition, the particle's angular momentum is constrained to be real and finite, thereby allowing a physically viable circular orbit. Failure of this condition indicates the absence of a physical timelike circular orbit at that radius. The limiting radius of the unstable circular orbit is determined by the condition \(\mathcal{A}_1(r_0)=0\).
Physical timelike orbits are therefore obtained by perturbing away from the limiting radius $r_{0+}$ as 
\begin{equation}
r_0 = r_{0+} + \epsilon\,,
\end{equation}
where $\epsilon>0$  merely ensures that the strict inequality \eqref{ineq} is satisfied and does not introduce any additional physical scale. 

In addition, circular timelike orbits are defined only in the domain of outer communication,
$r_0>r_+$, where the Killing vector $\partial_t$ remains timelike (see, e.g., Ref.~\cite{chandrasekhar1998mathematical}). The same argument applies to null orbits as well. These conditions will be imposed throughout the numerical analysis. Moreover, since we restrict attention to equatorial circular motion, we set $\theta = \pi/2$ without loss of generality.
\subsection{Dynamics of null-like geodesics}
\label{s4}
Unstable circular null geodesics play a central role in determining black hole shadows and related optical phenomena, which are directly relevant for astrophysical observations. Consequently, the instability of these orbits provides a natural connection between chaotic dynamics and observable properties of black holes. Since the normalization condition for null geodesics differs from that of timelike geodesics, the corresponding Lyapunov exponent also takes a different form. In this section, we briefly derive the Lyapunov exponent of circular null orbits by following the standard procedure presented in Refs.~\cite{touati2024lyapunov, kumara2024lyapunov, guo2022probing}. The Lagrangian \eqref{lang} and Hamiltonian \eqref{h1} equations of motion remain applicable in this case. 

For null geodesics, the conserved energy and angular momentum associated with the Killing vectors $\partial_t$ and $\partial_\phi$ are given by
\begin{equation}
\label{consnull}
E = h(r)\,\dot{t},
\qquad
L = D(r)\,\dot{\phi},
\end{equation}
from which one immediately obtains
\begin{equation}
\label{tdot}
\dot{t}=\frac{E}{h(r)}\,.
\end{equation}
Equation~\eqref{tdot} reflects the gravitational influence of the temporal metric component $h(r)$ on null propagation. In particular, the factor $E/h(r)$ exhibits a structure analogous to that associated with gravitational time dilation, while the locally measured photon frequency in gravitational redshift phenomena scales as $E/\sqrt{h(r)}$ \cite{radosz2013nature}. Since both effects originate from the same temporal metric component, Eq.~\eqref{tdot} will play an important role in the subsequent analysis of the Lyapunov exponent. These aspects will be discussed in greater detail in the subsequent sections. By using the Lagrangian equation \eqref{lang} and Eq. \eqref{consnull} together with the normalization condition for null-like orbits, \(ds^2=0\), the radial equation of motion can be written as
\begin{equation*}
\dot{r}^{2}=f\left(\frac{E^{2}}{h}-\frac{L^{2}}{D}\right)\,.
\end{equation*}
It is convenient to recast this equation into a mechanical form by introducing the effective potential
\begin{equation*}
V_{\rm eff}(r)\equiv f\left(\frac{L^{2}}{D}-\frac{E^{2}}{h}\right)\,,
\end{equation*}
such that the radial motion satisfies
\begin{equation*}
\dot{r}^{2}+V_{\rm eff}(r)=0\,.
\end{equation*}
Circular null orbits (photon spheres) at \(r=r_0\) are determined by the conditions
\begin{equation}
\label{vefs}
V_{\rm eff}(r_0)=0\,, \qquad V_{\rm eff}'(r_0)=0\,.
\end{equation}
The first condition yields the ratio of the conserved quantities along null geodesics, defining the impact parameter $b_0 \equiv L/E$. 
For a circular null orbit located at $r=r_0$, this quantity satisfies
\begin{equation}
\label{bo}
\frac{1}{b_0^2}
=
\frac{E^2}{L^2}
=
\frac{h(r_0)}{D(r_0)}\,.
\end{equation}
By using Eq.~\eqref{bo}, the second condition in Eq.~\eqref{vefs} reduces to an equivalent condition stated in Eq.~\eqref{ineq} for timelike geodesics, which in the present case reads
\begin{equation}
\label{ps}
 h(r_0) D'(r_0) - h'(r_0) D(r_0) = 0\,,
\end{equation}
with the photon Lyapunov exponent derived in terms of the effective potential as
\begin{equation}
\label{lyapgeneral}
\lambda^{2}
=
-\frac{V_{\rm eff}''(r_{0})}{2\dot{t}^{2}}\,.
\end{equation}
By using Eq.~\eqref{tdot} in Eq.~\eqref{lyapgeneral},  we obtain the photon Lyapunov exponent as 
\begin{equation}
\lambda^{2}
=
-\frac{h^{2}(r_{0})}{2E^{2}}
\,V_{\rm eff}''(r_{0})
=
f(r_0)\left(
\frac{h(r_0)}{D(r_0)}
-\frac{1}{2}\,h''(r_0)
\right)
\, .
\label{lyapgeneral2}
\end{equation}
Equation \eqref{lyapgeneral2} represents the Lyapunov exponent associated with null circular orbits, evaluated at the photon sphere. Notably, this expression is independent of angular momentum. The expression derived here is general for static, spherically symmetric spacetimes, where the instability is governed by the curvature of the effective radial potential. A similar procedure can be extended to rotating geometries, with appropriate modifications to account for frame dragging and axial symmetry. 

\section{Shift-symmetric black hole solution in Horndeski gravity}
\label{sec2}
In this section, we consider a four-dimensional Einstein-Horndeski theory minimally coupled to a U(1) gauge field to identify the origin of chaos bound violations beyond the realm of GR. The Horndeski theory is the most general scalar–tensor framework yielding second-order field equations, thereby permitting higher-derivative interactions without introducing Ostrogradsky instability. Within this framework, black hole solutions have been widely studied in Refs.~\cite{h1,h3,h4,h5, santos2023ads, santos2024magnetized}. Building upon these results, the present work investigates possible departures from the MSS bound and their geometric origin in scalar-tensor gravity. This provides a suitable setting for exploring how modifications of the spacetime geometry induced by the scalar sector may affect chaotic behavior near black holes. In accordance with the framework developed in Refs.~\cite{horndeski1974second, anabalon2014asymptotically, babichev2017asymptotically, badia2017gravitational, carvajal2025study, myung2025extended, liang2025einstein}, we write the action of the Horndeski theory as
\begin{equation}
\label{eq:action}
\mathcal{S}=\int d^4x \,\sqrt{-g}\Bigl[R - 2\Lambda - \frac{1}{4}F_{\mu\nu}F^{\mu\nu}
-\frac{1}{2}\bigl(\alpha\,g_{\mu\nu} - \zeta\,G_{\mu\nu}\bigr)\nabla^\mu\varphi\,\nabla^\nu\varphi\Bigr]\,,
\end{equation}
where $F=dA$ is the ${\rm U}(1)$ gauge field strength, $\varphi$ is a real scalar field, $\alpha$ is a dimensionless coupling constant, and $\zeta$ is a coupling constant of dimension [length]$^2$.  The scalar enters only through its derivatives, so the action is invariant under the shift $\varphi\mapsto\varphi+\mathrm{constant}$, implying a conserved current. Variation of Eq. \eqref{eq:action} leads to the field equations 
\begin{eqnarray}
\label{eq:Einstein}
G_{\mu\nu} + \Lambda g_{\mu\nu} &=& \frac{1}{2}\left[\alpha\,T_{\mu\nu} + \zeta\,\xi_{\mu\nu} 
+ \left(\frac{1}{2}F_{\mu\nu}^2 - \frac{1}{8}F^2\,g_{\mu\nu}\right)\right]\,,\\
\label{eq:scalar}
\nabla_\mu\Bigl[(\alpha\,g^{\mu\nu}-\zeta\,G^{\mu\nu})\nabla_\nu\varphi\Bigr] &=& 0\,,\\
\label{eq:maxwell}
\nabla_\nu F^{\nu\mu} &=& 0\,.
\end{eqnarray}
The tensor $T_{\mu\nu}$ is the canonical energy-momentum of the scalar and $\xi_{\mu\nu}$ arises from the nonminimal coupling. Explicitly, one finds 
\begin{align}
T_{\mu\nu} &= \nabla_\mu\varphi\,\nabla_\nu\varphi - \frac{1}{2} g_{\mu\nu}\nabla_\rho\varphi\nabla^\rho\varphi\,,\\
\xi_{\mu\nu} &= \frac{1}{2}\nabla_\mu\varphi\,\nabla_\nu\varphi\,R - 2\nabla_\rho\varphi\,\nabla_{(\mu}\varphi\,R^\rho_{\nu)} 
- \nabla^\rho\varphi\,\nabla^\lambda\varphi\,R_{\mu\rho\nu\lambda} \notag\\
&\quad - (\nabla_\mu\nabla^\rho\varphi)(\nabla_\nu\nabla_\rho\varphi) + (\nabla_\mu\nabla_\nu\varphi)\,\Box\varphi 
+ \frac{1}{2}G_{\mu\nu}(\nabla\varphi)^2 \notag\\
&\quad - g_{\mu\nu}\Bigl[-\frac{1}{2}(\nabla^\rho\nabla^\lambda\varphi)(\nabla_\rho\nabla_\lambda\varphi) 
+ \frac{1}{2}(\Box\varphi)^2 - \nabla_\rho\varphi\,\nabla_\lambda\varphi\,R^{\rho\lambda}\Bigr]\,.
\end{align}
From the scalar field equation \eqref{eq:scalar}, one can define the current 
\begin{equation}
J^\mu = (\alpha\,g^{\mu\nu} - \zeta\,G^{\mu\nu})\nabla_\nu\varphi\,,
\end{equation}
which is covariantly conserved and is used to demonstrate a no-hair theorem in this theory \cite{hui2013no}. The corresponding static spherically symmetric black hole solutions can be derived from the metric and field ansatz given in Eq. \eqref{hb}, with
\begin{equation}
A = \Phi_{\rm t}(r)\,dt\,,
\end{equation}
where $\Phi(r)$ is the ${\rm U}(1)$  gauge field potential with $\Phi_{\rm t}(r)$ as the only non-vanishing component, $d\Omega^2$ is the line element on the unit 2-sphere.  By inserting this ansatz into the Maxwell equation \eqref{eq:maxwell}, we get
\begin{equation}
\label{eq:Phi_prime}
\Phi' = \frac{4Q}{r^2}\sqrt{\frac{h}{f}}\,,
\end{equation}
where \(Q\) represents the integration constant identified with the electric charge of the black hole.  With this result, the remaining field equations reduce to a set of coupled ordinary differential equations for $h$, $f$, and $\Phi$. The resulting exact black hole solution has the form \cite{cisterna2014asymptotically, liang2025einstein}
\begin{eqnarray}
h &=& 1 - \frac{2M}{r} + \frac{4Q^2}{r^2} - \frac{4Q^4}{3r^4}\,, \label{h0}\\
f &=& \frac{r^4}{(r^2 - 2Q^2)^2}\,h,\label{h2}\\
\Phi &= &\Phi_\infty - \frac{4Q}{r} + \frac{8Q^3}{3r^3}\,,\label{h3}\\
\varphi'(r) &=& \sqrt{-\frac{8Q^2}{\zeta\,r^2f}}\,, \label{scala}
\end{eqnarray}
where $M$ denotes the integration constant associated with the black hole mass. Notably, these corrections significantly modify the near-horizon region and give rise to a new curvature singularity at a finite radius, thereby impacting the dynamical properties of the particle trajectories.  We fix the gauge by setting the electric potential to vanish at infinity ($\Phi_\infty=0$). The reality condition $\zeta Q^2<0$ fixes the sign of the Horndeski coupling to $\zeta<0$. We emphasize that the perturbative stability of hairy black holes in shift-symmetric Horndeski theories is a separate issue and has been shown to depend on the specific solution and perturbation sector \cite{ogawa2016instability,takahashi2017linear}. Here we restrict our analysis to the classical stability of circular geodesics on the fixed black hole background and the linear perturbative stability of the background itself is beyond the scope of this work.

Moreover, this spacetime exhibits curvature singularities at $r=0$ and at $r_*=\sqrt{2}Q$, where the Kretschmann scalar $K=R_{\mu\nu\rho\sigma}R^{\mu\nu\rho\sigma}$ diverges. In accordance with the cosmic censorship conjecture, the avoidance of a naked singularity requires that the event horizon shield all curvature singularities. For the charged Horndeski black hole, this condition translates into the bound
\begin{equation}
\frac{Q}{M} < \frac{3\sqrt{2}}{8}\,,
\end{equation}
that ensures both the existence of a real event horizon and its location outside the curvature singularity at $r_{*}=\sqrt{2}Q$, thereby yielding a well-defined charged Horndeski black hole solution~\cite{cisterna2014asymptotically}.

Recently, this black hole was shown to undergo an ultra-slow Hawking evaporation process \cite{liang2025einstein}, highlighting its dynamical longevity and reinforcing its relevance as a physically viable background. Motivated by these developments, we investigate the dynamical stability and chaotic properties of this spacetime by analyzing the Lyapunov growth of both timelike and null geodesic orbits, with particular emphasis on the behavior of the MSS chaos bound.

\subsection{Chaotic regimes in Horndeski gravity}
\label{s3}
The analysis of chaotic dynamics basically relies on determining circular orbits. While this is usually a non-trivial task, in the present case, we found that timelike circular orbits satisfy the polynomial condition
\begin{equation}
\label{r0}
r_0^3 (r_0 - 3M) + 8Q^2 r_0^2 - 4Q^4 > 0\, ,
\end{equation}
which follows from the inequality condition given in Eq. \eqref{ineq}. Although the orbiting particle may carry an electric charge, we find that this charge does not affect the radial location of the circular orbit. Instead, it enters the dynamics through the angular momentum, thereby modifying the orbital velocity without shifting the orbital radius. This result shows that the circular orbits are determined entirely by the spacetime geometry and are independent of the characteristics of the test particles. In the limiting case where the inequality \eqref{ineq} is saturated, Eq.~\eqref{r0} reduces to the condition for unstable circular null geodesics in the Horndeski black hole background. Consequently, for timelike circular orbits, the inequality must be satisfied strictly; otherwise, the angular momentum becomes ill-defined, and the orbit ceases to be physical. This highlights that well-defined equilibrium trajectories are restricted to a particular region of spacetime.

The event horizon of the black hole is located at the largest root of
\(
h(r_+)=0\,.
\)
For a static, spherically symmetric spacetime with metric functions \(f\) and \(h\), the surface gravity is given by
\begin{equation}
\kappa
=
\frac{1}{2}
\left.
\sqrt{\frac{h(r)}{f(r)}}\, f'(r)
\right|_{r=r_+}\,,
\label{SGG}
\end{equation}
which defines the characteristic thermodynamic scale corresponding to the black hole temperature. In order to examine the chaos bound, we compare the Lyapunov exponent
associated with radial perturbations of circular orbits with the surface gravity of the horizon. We characterize a violation through
\begin{equation}
\lambda^2-\kappa^2>0,
\end{equation}
which, whenever $\kappa\neq0$, is equivalently expressed as $\lambda/\kappa>1$. The former criterion is more convenient in the near-extremal regime, where both $\lambda$ and $\kappa$ may vanish and the ratio $\lambda/\kappa$ can otherwise take the indeterminate form. In such cases, however, the simultaneous vanishing $\lambda=\kappa=0$ corresponds to saturation of the bound in the quadratic sense, $\lambda^2-\kappa^2=0$, rather than to an ambiguity in the chaos-bound criterion. We therefore use both $\lambda^2-\kappa^2$ and $\lambda/\kappa$ throughout the analysis, with the latter employed only when $\kappa\neq0$, to make the onset and saturation of the bound unambiguous. In the following, we investigate the parameter regimes under which the shift-symmetric Horndeski black hole admits departures from the MSS bound.

\begin{table*}[t]
\centering
\caption{Chaos-bound behavior in Horndeski gravity for different parameter choices. The geometric quantities are identical across cases, while $L_0$ and $\lambda^2-\kappa^2$ depend on $\epsilon$ and $q$.}
\label{Thon}
\renewcommand{\arraystretch}{1.12}
\setlength{\tabcolsep}{6pt}
\small
\begin{tabular}{c c c c @{\hspace{0.6cm}} c c @{\hspace{0.6cm}} c c}
\hline\hline
& & & &
\multicolumn{2}{c}{$\epsilon = 10^{-5},\,q=0.99$} &
\multicolumn{2}{c}{$\epsilon = 10^{-3},\,q=0.99$} \\
\cline{1-8}
$\dfrac{Q}{M}$ &
$\Delta\!\left(\dfrac{Q}{M}\right)$ &
$r_{+}$ &
$r_{0}$ &
$L_{0}$ &
$\lambda^{2}-\kappa^{2}$ &
$L_{0}$ &
$\lambda^{2}-\kappa^{2}$ \\
\hline
0.529 & $-0.00133009$ & 0.894535 & 1.82460 & 370378.43 & 0.0130599 & 3707.07 & 0.0130425 \\
0.528 & $-0.00233009$ & 0.928098 & 1.83478 & 367320.56 & 0.00546309 & 3676.53 & 0.00544470 \\
0.527 & $-0.00333009$ & 0.953790 & 1.84472 & 364455.12 & 0.000154448 & 3647.90 & 0.000135096 \\
0.526 & $-0.00433009$ & 0.975225 & 1.85444 & 361761.86 & $-0.00389605$ & 3620.99 & $-0.00391630$ \\
0.500 & $-0.03033009$ & 1.23394 & 2.05600 & 321219.06 & $-0.0292060$ & 3215.86 & $-0.0292370$ \\
0.400 & $-0.13033009$ & 1.61086 & 2.49321 & 255777.44 & $-0.0310386$ & 2561.92 & $-0.0310702$ \\
0.300 & $-0.23033009$ & 1.80208 & 2.73868 & 197564.29 & $-0.0284328$ & 1980.64 & $-0.0284611$ \\
0.200 & $-0.33033009$ & 1.91683 & 2.88952 & 134740.53 & $-0.0267065$ & 1354.32 & $-0.0267327$ \\
0.000 & $-0.53033009$ & 2.00000 & 3.00000 & 948.686 & $-0.0254636$ & 94.9000 & $-0.0255246$ \\
\hline\hline
\end{tabular}
\end{table*}

\begin{table*}[t]
\centering
\caption{Chaos-bound behavior for neutral timelike circular orbits ($\epsilon=0.5$) and null circular orbits (photon sphere) in the charged Horndeski black hole. The geometric quantities are identical, while the dynamical indicators differ.}
\label{neutral_photon}
\renewcommand{\arraystretch}{1.12}
\setlength{\tabcolsep}{2.5pt}
\small
\resizebox{\textwidth}{!}{%
\begin{tabular}{c c c @{\hspace{0.40cm}} c c @{\hspace{0.40cm}} c c @{\hspace{0.40cm}} c}
\hline\hline
& & &
\multicolumn{2}{c}{Neutral orbits ($\epsilon=0.5$)} &
\multicolumn{2}{c}{$q=0.99,\ \epsilon=0.5$} &
Photon orbits \\
\cline{1-8}
$\dfrac{Q}{M}$ &
$r_{+}$ &
$r_{0}$ &
$L_{0}$ &
$\lambda^{2}-\kappa^{2}$ &
$L_{0}$ &
$\lambda^{2}-\kappa^{2}$ &
$\lambda^{2}-\kappa^{2}$ \\
\hline
0.529 & 0.894535 & 1.82460 & 3.31753 & $-0.0122852$ & 10.9554 & 0.00120799 & 0.0130600 \\
0.528 & 0.928098 & 1.83478 & 3.32505 & $-0.0201276$ & 10.9163 & $-0.00668122$ & 0.00546328 \\
0.527 & 0.953790 & 1.84472 & 3.33261 & $-0.0256624$ & 10.8791 & $-0.0122632$ & 0.000154643 \\
0.526 & 0.975225 & 1.85444 & 3.34020 & $-0.0299214$ & 10.8437 & $-0.0165697$ & $-0.00389584$ \\
0.500 & 1.23394 & 2.05600 & 3.53490 & $-0.0573596$ & 10.2578 & $-0.0452289$ & $-0.0292057$ \\
0.400 & 1.61086 & 2.49321 & 4.11641 & $-0.0567799$ & 9.16787 & $-0.0480696$ & $-0.0310383$ \\
0.300 & 1.80208 & 2.73868 & 4.50339 & $-0.0515894$ & 8.18443 & $-0.0452258$ & $-0.0284325$ \\
0.200 & 1.91683 & 2.88952 & 4.75714 & $-0.0482411$ & 7.14055 & $-0.0438983$ & $-0.0267063$ \\
0.000 & 2.00000 & 3.00000 & 4.94975 & $-0.0458403$ & 4.94975 & $-0.0458403$ & $-0.0254630$ \\
\hline\hline
\end{tabular}%
}
\end{table*}

In Tables~\ref{Thon} and \ref{neutral_photon}, we present representative examples in which the shift-symmetric Horndeski black hole exhibits apparent violations of the MSS chaos bound. Throughout, we restrict to the physically admissible regime in which the event horizon encloses the curvature singularity at $r_*=\sqrt{2}Q$, enforced by
\begin{equation}
\Delta\!\left(\frac{Q}{M}\right)
= \frac{Q}{M}-\frac{3\sqrt{2}}{8}<0\,,
\end{equation}
thereby ensuring consistency with cosmic censorship.

We investigate unstable circular orbits at varying distances from the event horizon, parameterized by $\epsilon$, where smaller values correspond to trajectories located closer to $r_+$. As shown in Table~\ref{Thon}, apparent departures from the MSS bound occur for charged test particles situated near the event horizon, yet still outside the photon sphere (e.g., $\epsilon=10^{-3},\,10^{-5}$), as the charge-to-mass ratio approaches the naked singularity threshold. 
In contrast, Table~\ref{neutral_photon} demonstrates that for uncharged particles located farther from the black hole ($\epsilon=0.5$), no violations are observed, even in regimes nearing the naked singularity. This indicates that the emergence of chaos-bound violations is governed by two key factors: proximity to the naked singularity and the radial location of the orbit relative to the event horizon. In particular, in the near-singular regime ($Q/M \approx 0.527~\text{to}~0.529$), all photon orbits exhibit chaotic behavior. This can be attributed to their inherently closer proximity to the horizon, combined with the enhanced influence of the near-singular geometry.

It is also important to isolate the role of the particle’s electric charge. The charge of the orbiting massive particle couples to the electromagnetic field of the black hole, thereby modifying its dynamics through the angular momentum. As seen in Table~\ref{neutral_photon}, even for orbits located relatively far from the black hole (e.g., $\epsilon=0.5$), the presence of charge leads to an increase in the angular momentum. As the system approaches the near-singular regime, this effect contributes to the onset of chaos-bound violations. A robust trend emerges: violations become increasingly pronounced as the curvature singularity approaches the event horizon, namely as
\begin{equation}
\frac{Q}{M} \longrightarrow \frac{3\sqrt{2}}{8}\,.
\end{equation}

This behavior indicates a geometric origin of the violation. As the charge-to-mass ratio increases, the curvature singularity moves toward the event horizon, pushing the system into a regime where the thermodynamic scale collapses before a naked singularity is reached. In contrast, the unstable circular orbit remains outside the horizon and continues to generate a non-vanishing Lyapunov exponent. This mismatch between the suppressed surface gravity and the persistent orbital instability leads directly to $\lambda > \kappa$. Our central result is that the shift-symmetric Horndeski black hole provides a fully classic example in which the Lyapunov exponent associated with unstable circular orbits exceeds the surface gravity, despite the absence of naked singularities. While cosmic censorship imposes the bound $Q/M < 3\sqrt{2}/8$, our numerical analysis indicates that chaos-bound violations are absent below a more restrictive threshold
$Q/M\leq0.526$ as inferred empirically from Tables~\ref{Thon}--\ref{neutral_photon}.  Consistent with the behavior observed for timelike orbits, we find that the chaos bound is violated for null geodesics in the near-singular regime, even prior to encountering the naked singularity, as explicitly demonstrated in Table~\ref{neutral_photon}. This confirms that the phenomenon is generic with respect to the probe and is intrinsically associated with the geometry of the spacetime. 
\section{Chaos bound violations for a 4D Einstein-Gauss-Bonnet gravity}
\label{sec3}
In this section, to further clarify whether the chaos-bound violations emanate from higher-order curvature modifications, as we have suspected in our previous study \cite{Targema2025MSS}, we now turn our attention to EGB gravity. This theory incorporates a quadratic curvature correction through the GB  term, governed by the coupling constant $\alpha$, thereby offering a suitable model to investigate whether these corrections alone can lead to violation of the MSS bound. Therefore, we examine a 4D static and spherically symmetric charged black hole solution in EGB theory, for which the gravitational action is defined in an arbitrary dimension $D$ as follows \cite{glavan2020einstein}:
\begin{equation}
\mathcal{S}
= \frac{1}{16\pi}
\int d^D x \,
\left(
R
+ \frac{\alpha}{D-4}\mathbf{L_{GB}}
- F_{\mu\nu}F^{\mu\nu}
\right)\,,
\end{equation}
where
\[
\mathbf{L_{GB}}
= R_{\mu\nu\rho\sigma}R^{\mu\nu\rho\sigma}
- 4 R_{\mu\nu}R^{\mu\nu}
+ R^{2}
\]
is the GB invariant, and $\alpha$ denotes the GB coupling constant controlling higher–curvature corrections. The factor \((D-4)^{-1}\) realizes the regularized 4D limit of the theory \cite{cognola2013einstein, yang2020weak}.  Notably, the dimensional-regularization prescription underlying the original 4D EGB construction has been subject to debate, particularly regarding the existence of a well-defined 4D theory at the level of the metric field equations and the additional degrees of freedom that arise in consistent formulations \cite{gurses2020there, fernandes20224d, mahapatra2020note}. However, our analysis is confined to the static, spherically symmetric sector, for which the black hole solution
is also obtained in consistent regularized formulations of a 4D EGB gravity, so that our conclusions concerns this highly symmetric black hole sector and do not rely on the disputed interpretation of the 4D theory as a purely metric theory. 

The electromagnetic sector preserves the same Maxwell structure encountered in the Horndeski black hole discussed earlier. In this case, the spacetime is sourced by the gauge four-potential
\begin{equation}
A_{\mu}\,dx^{\mu}
=
\frac{Q}{r}\,dt.
\end{equation}

The static and spherically symmetric 4D EGB black hole solution introduced in Ref.~\cite{glavan2020einstein} is described by the line element with the following metric function:
\begin{equation}
f(r)=h(r)
= 1
+ \frac{r^{2}}{2\alpha}
\left[
1
- \sqrt{
1
+ 4\alpha
\left(
\frac{2M}{r^{3}}
- \frac{Q^{2}}{r^{4}}
\right)
}
\right]\,.
\end{equation}
Here $M$ and $Q$ represent the black hole mass and electric charge, respectively. The GB coupling $\alpha$ governs how the theory departs from GR. However, our analysis reveals that its effect on chaos-bound violations is not fundamental; rather, it lies in modifying how the geometry approaches extremality. Furthermore, in the limit $\alpha \to 0$, the solution smoothly reduces to the standard Reissner–Nordstr\"om geometry.

The event horizon is determined by the largest root of the condition $f(r)=0$, leading to the following expression:
\begin{equation}
r^{2}- 2Mr+ Q^{2}+ \alpha= 0\,,
\end{equation}
whose solutions yield the inner and outer horizon radii
\begin{equation}
\label{ehor}
r_{\pm}= M\pm\sqrt{M^{2} - Q^{2} - \alpha}\,.
\end{equation}
A regular black hole horizon exists at $r_{+}$ provided the parameter space satisfies
\begin{equation}
\label{ee1}
M^{2} \geq Q^{2} + \alpha\,.
\end{equation}
The extremality condition shows that the degeneracy of the horizon is governed by the combined contribution $Q^{2}+\alpha$, rather than on $Q$ and $\alpha$ separately. This indicates that an effective charge parameter governs the transition to extremality, determining the near-horizon configuration and the corresponding instability scales. For example, the equality part of Eq. \eqref{ee1} corresponds to the extremal configuration, where the two horizons coincide, while violation of this bound results in the emergence of a naked singularity. Consequently, modifying either $Q$ or $\alpha$ similarly influences the geometry as it approaches extremality. 

From the standpoint of geometric and energetic consistency, the GB coupling is typically restricted to non-negative values, $\alpha \ge 0$. When the higher-curvature contributions are recast as an effective matter sector in the Einstein frame, a positive GB coupling preserves the weak energy condition, thereby maintaining standard attractive gravitational behavior and ensuring the validity of several geometric inequalities governing null orbits and horizon structure. By contrast, $\alpha<0$ generically induces violations of the weak energy condition, leading to qualitatively altered causal and optical properties, including reversed photon-sphere/shadow bounds and the potential emergence of pathological horizon configurations \cite{guo2020innermost}. To avoid such energetically exotic regimes, we confine our analysis to $\alpha \ge 0$, where the black hole geometry remains physically admissible and the horizon structure is well defined across the allowed parameter space.
In what follows, we investigate how curvature modifications induced by GB corrections influence the chaos bound and its possible violations. Notably, black holes in this class of theories have been studied extensively across a broad range of physical contexts. For instance, within string theory—particularly from the perspective of low–energy effective actions—the GB coupling encodes leading-order quantum-gravity corrections and is closely related to the string scale \cite{gross1987quartic, charmousis2002general}.

In the context of the MSS conjecture, the chaotic dynamics of massive gravitons in this family of solutions were analyzed in Ref.~\cite{fayyaz2025chaotic}, where chaotic behavior was attributed to the graviton mass and the GB coupling. Likewise, Ref.~\cite{xie2023circular} reported that the combined effects of the GB term and angular momentum can drive the system into regimes where the MSS chaos bound may be violated.

In the present study, we revisit this issue from a complementary perspective. Our analysis suggests that, although chaos–bound violations can apparently arise in this spacetime, their origin is fundamentally geometric rather than being directly driven by the coupling parameter or by angular momentum itself. In particular, we argue that previously reported violations are associated with effectively unconstrained orbital angular momentum. In this sense, the GB coupling and angular momentum act as facilitators rather than primary sources of the violation, whose underlying mechanism remains rooted in the near–extremal geometry.

\subsection{Chaotic regions of the 4D Einstein-Gauss-Bonnet black hole spacetime}
In analyzing the chaos bound in this spacetime, we first isolate the role of the GB coupling $\alpha$.  Insights follows directly from the event horizon and extremal conditions, which we have already discussed in Eqs. (\ref{ehor}) and \eqref{ee1}. Moreover, since both the charge, mass and the GB parameter are already bounded by cosmic censorship, varying $Q$ at fixed $\alpha$, or varying $\alpha$ at fixed $Q$, is physically equivalent so long as the analysis remains within the black hole sector. What matters is the approach to extremality, since it is this limit that determines the near-horizon redshift structure and the instability scales relevant for chaos growth.
For definiteness, we fix $\alpha=0.2$ and vary $Q$, ensuring that all configurations considered satisfy the horizon condition. An analogous analysis could be performed by fixing $Q$ and varying $\alpha$, leading to identical conclusions, as only the bounded combination $Q^2+\alpha$ regulates the parametric approach to extremality and potential bound violation.
In Table \ref{tab:GB_master}, we demonstrate that the chaos bound is apparently violated in this spacetime for charged and neutral probes, encompassing both timelike and null circular orbits. The roles of the orbital parameter $\epsilon$ and the test charge follow the same qualitative trends observed previously in the Horndeski black hole analysis. 

Notably, the violations persist at fixed $\alpha$, indicating that the GB coupling itself is not the primary driver of the effect, in contrast to the interpretation put forward in Ref.~\cite{fayyaz2025chaotic}. This indicates that the violation is not caused by the higher-order curvature correction but reflects a mechanism inherently present in GR. A familiar pattern emerges: the onset and strengthening of violations correlate with the charge-to-mass ratio approaching extremality. This provides strong evidence that the apparent violations of the MSS bound arise from the geometric properties of near-extremal black holes, where diminishing surface gravity coexists with the persistence of unstable circular orbits. For the GB black hole, this behavior is fully consistent with the results of \cite{xie2023circular}. The key distinction, however, is that the coupling $\alpha$ remains effectively inert in triggering the violation; even in the limit $\alpha \to 0$, increasing charge in the Reissner-Nordstr\"om sector drives the geometry toward extremality, with the violation persisting in this limit (see Table \ref{tab:RN-chaos}). In the following section, we therefore turn to the geometric origin underlying these violations.

\begin{table*}[t]
\centering
\caption{Chaos-bound behavior for circular orbits in the 4D EGB black hole spacetime with $\alpha=0.2$, and the test charge $q=1.0$. The geometric quantities are identical across all cases, while the dynamical quantities depend on the charge and the orbital parameter $\epsilon$.}
\label{tab:GB_master}

\renewcommand{\arraystretch}{1.10}
\setlength{\tabcolsep}{2.5pt}
\small

\resizebox{\textwidth}{!}{%
\begin{tabular}{
c c c
@{\hspace{0.18cm}}
c c
@{\hspace{0.18cm}}
c c
@{\hspace{0.18cm}}
c c
@{\hspace{0.18cm}}
c
}
\hline\hline

&
&
&
\multicolumn{2}{c}{$\epsilon=10^{-5}$}
&
\multicolumn{2}{c}{$\epsilon=10^{-3}$}
&
\multicolumn{2}{c}{$\epsilon=0.50$}
&
Null
\\

\cline{1-10}

$Q/M$ &
$r_{+}$ &
$r_{0}$ &
$L_{0}$ &
$\lambda_{t}^{2}-\kappa^{2}$ &
$L_{0}$ &
$\lambda_{t}^{2}-\kappa^{2}$ &
$L_{0}$ &
$\lambda_{t}^{2}-\kappa^{2}$ &
$\lambda_{\mathrm{null}}^{2}-\kappa^{2}$
\\

\hline

0.85 & 1.27839 & 2.21521 &
170984. & 0.0121161 &
1713.66 & 0.0120570 &
6.84235 & $-0.0120214$ &
0.0121167 \\

0.80 & 1.40000 & 2.32524 &
154382. & 0.00385967 &
1548.21 & 0.00379519 &
6.70184 & $-0.0209724$ &
0.00386032 \\

0.60 & 1.66332 & 2.61855 &
109880. & $-0.00911465$ &
1105.68 & $-0.00918376$ &
6.30649 & $-0.0331574$ &
$-0.00911395$ \\

0.40 & 1.80000 & 2.78686 &
72704.9 & $-0.0133919$ &
738.137 & $-0.0134598$ &
5.90362 & $-0.0356078$ &
$-0.0133912$ \\

0.20 & 1.87178 & 2.87782 &
36372.7 & $-0.0151056$ &
385.845 & $-0.0151718$ &
5.43433 & $-0.0355250$ &
$-0.0151050$ \\

0.00 & 1.89443 & 2.90681 &
934.328 & $-0.0155802$ &
93.4646 & $-0.0156348$ &
4.89356 & $-0.0342377$ &
$-0.0155797$ \\

\hline\hline
\end{tabular}%
}
\end{table*}
\section{Geometric origin of chaos-bound violations in general relativity}
\label{sec4}
Since chaos-bound violations have been shown to arise independently of particle dynamics and are not tied to any particular modification of GR, we now aim to narrow down our focus to identify their fundamental origin. In this section, we demonstrate that these violations are driven by a universal geometric mechanism associated with the near-horizon structure of black hole spacetimes. In particular, our analysis reveals that the apparent violation of the MSS chaos bound originates from the decoupling between the thermodynamic scale, set by the surface gravity, and the dynamical instability scale, governed by the curvature of the effective potential. We show that this behavior is controlled by the gravitational time dilation experienced by unstable circular orbits near the horizon: when the orbit approaches the degenerate horizon, the associated infinite redshift suppresses the instability scale, whereas orbits that remain separated from the horizon continue to produce a nonvanishing Lyapunov exponent.  
\subsection{The Reissner-Nordstr\"om black hole}
We now examine the limit $\alpha=0$, in which the EGB solution reduces to the Reissner–Nordstr\"om black hole. This provides a minimal setting in GR to isolate the geometric origin of the chaos-bound violation, free of higher-curvature or scalar-tensor modifications. For the Reissner–Nordstr\"om spacetime, the horizon radii are
\begin{equation}
\label{eqr}
r_{\pm}(Q,M)=M\pm\sqrt{M^{2}-Q^{2}} \, ,
\end{equation}
while the circular geodesics satisfy
\begin{eqnarray}
\label{eqr0}
r_{0}(Q,M)= \frac{3M+\sqrt{9M^{2}-8Q^{2}}}{2}\,,\\
r_{0}(Q,M)> \frac{3M+\sqrt{9M^{2}-8Q^{2}}}{2}\,,
\label{eqr01}
\end{eqnarray}
where the equality part (Eq. \eqref{eqr0}) represents null circular orbits, and the inequality part (Eq. \eqref{eqr01}) represents timelike circular orbits. It follows from these relations that the charge-to-mass ratio determines the structure of the horizon as well as the allowed circular orbit. Notably, as $Q\to M$, a degenerate horizon forms, playing a crucial role in shaping the chaos bound. Furthermore, cosmic censorship requires $M\ge Q$, with extremality at $M=Q$ where $r_+=r_-=M$. Physical circular orbits must lie outside the event horizon, $r_0>r_+$, so the relevant solution corresponds to the largest admissible root of Eq.~\eqref{eqr0}. It is worth noting that, in the extremal limit, the inner circular orbit approaches the degenerate horizon and is associated with a vanishing Lyapunov exponent. This behavior is not unique to the Reissner--Nordstr\"om spacetime but appears to arise in several geometries, including all static black hole solutions considered in this work. 

In the extremal limit, as the Reissner-Nordstr\"om black hole develops a degenerate horizon and the near-horizon region becomes $AdS_{2}\times S^{2}$, the surface gravity vanishes in this limit, reflecting the collapse of the thermodynamic scale associated with the horizon. If the chaos bound were universally tied to this scale, one would expect the Lyapunov exponent to vanish as well.
However, the unstable circular null orbit persists outside the degenerate horizon. In the extremal geometry, the only admissible photon orbit lies at $r_0=2M$, and its Lyapunov exponent evaluates to
\[
\lambda(r_0=2M)=\frac{1}{4\sqrt{2}\,M}\,,
\]
which remains nonzero even though the surface gravity is zero. Consequently, the chaos bound is violated because the instability scale does not vanish along with the thermodynamic scale. The near-horizon throat itself (in Poincar\'e coordinates) has the form given by \cite{pradhan2011circular}
\begin{equation}
ds^{2}=M^{2}\left(-r^{2}dt^{2}+\frac{dr^{2}}{r^{2}}+d\Omega^{2}\right)\,,
\label{rind}
\end{equation}
does not support genuinely unstable circular null geodesics. The reason is geometric; in this $AdS_{2}\times S^{2}$ region, the two-sphere has fixed areal radius $M$, independent of the radial coordinate of the $AdS_{2}$ factor. Thus, unlike the full black hole spacetime, the area function does not vary along the radial direction. 
As discussed earlier, circular geodesics arise from the extremization of the effective potential, which in spherically symmetric spacetimes is governed by the radial derivative of the areal radius. A necessary condition for a circular orbit is therefore a nontrivial radial variation of this geometric quantity. In the throat geometry, however, the areal radius is constant, so its radial derivative vanishes identically. Consequently, the extremization condition cannot be satisfied in a nontrivial way, and no isolated unstable circular geodesics exist within the pure $AdS_{2}\times S^{2}$ region (more details can be found in Ref. \cite{pradhan2011circular}). The finite, nonzero Lyapunov exponent observed at extremality therefore cannot originate from the throat itself but instead from the full spacetime outside the degenerate horizon, where the areal radius regains radial dependence and unstable orbits persist. 

The apparent violation originates from a geometric separation of scales that emerges as extremality is approached. Horizon degeneracy drives the surface gravity $\kappa$ to zero while simultaneously generating an increasingly long near-horizon throat. The collapse of this thermal scale, however, does not remove the dynamical instability of circular orbits that remain outside the throat. Since the corresponding Lyapunov exponent is determined by the local geometry at the unstable orbit rather than by the surface gravity itself, it generally remains finite. Consequently,
\(
    \lambda^2-\kappa^2>0,
\)
throughout the near-extremal regime.

This observation reveals a geometric tension in applying the MSS bound to black hole geodesics. Within its holographic interpretation, the instability probe is expected to remain synchronized with the near-horizon thermal scale. Near extremality, however, this synchronization can be lost because the unstable orbit no longer follows the geometry of the emerging AdS$_2$ throat, even though the thermodynamic scale is determined entirely by it. The resulting decoupling produces the apparent violation. Whether this reflects a genuine breakdown of the MSS bound or instead signals that the relevant instability must be analyzed directly within the near-horizon throat is the central question addressed in the remainder of this work. Before turning to that analysis, we first examine several representative black hole geometries in general relativity to determine whether nonsingular or rotating spacetimes exhibit qualitatively different behavior.
\begin{figure}[t]
\centering

\begin{minipage}{0.48\textwidth}
\centering
\includegraphics[width=\textwidth]{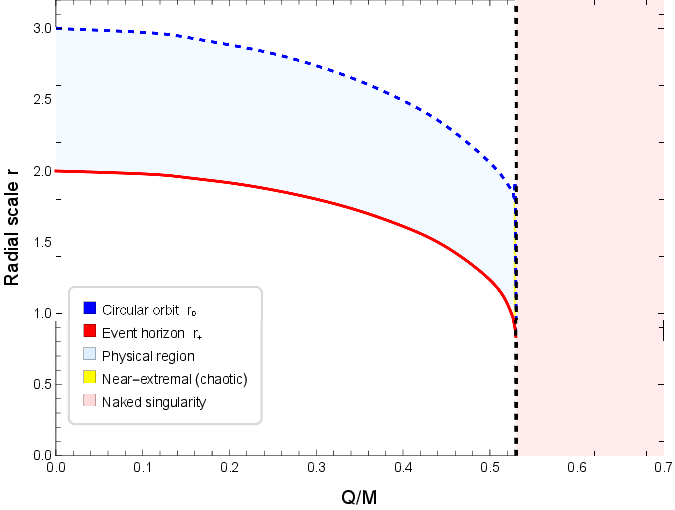}
\\ (a)
\end{minipage}
\hfill
\begin{minipage}{0.48\textwidth}
\centering
\includegraphics[width=\textwidth]{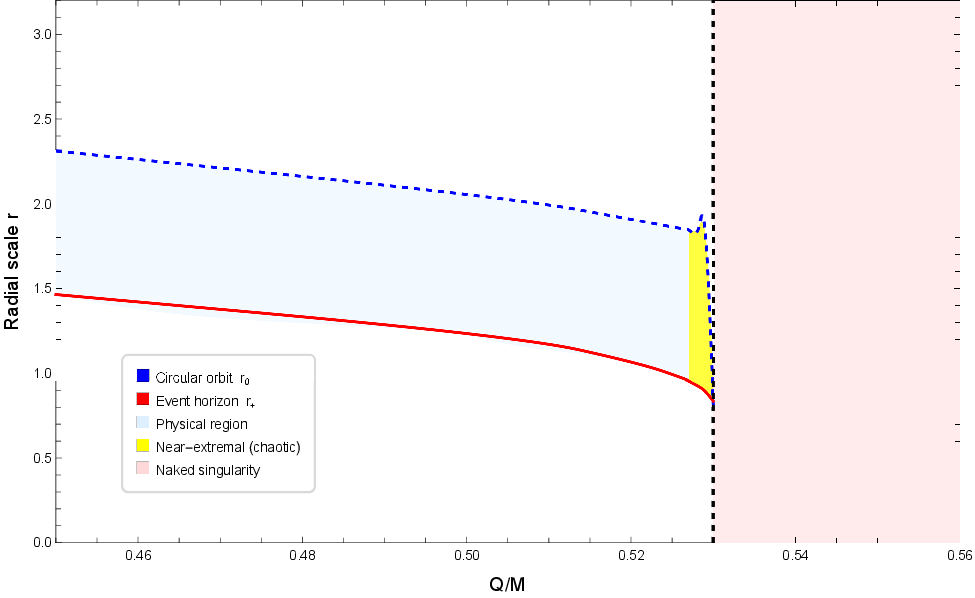}
\\ (b)
\end{minipage}
\hfill
\begin{minipage}{0.48\textwidth}
\centering
\includegraphics[width=\textwidth]{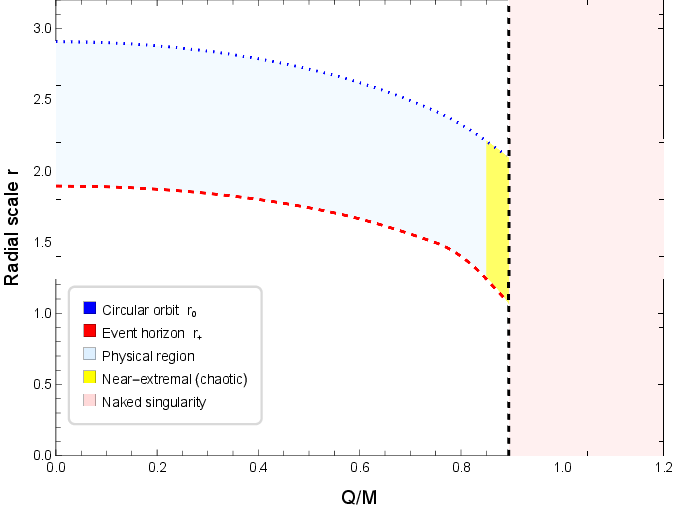}
\\ (c)
\end{minipage}
\hfill
\begin{minipage}{0.48\textwidth}
\centering
\includegraphics[width=\textwidth]{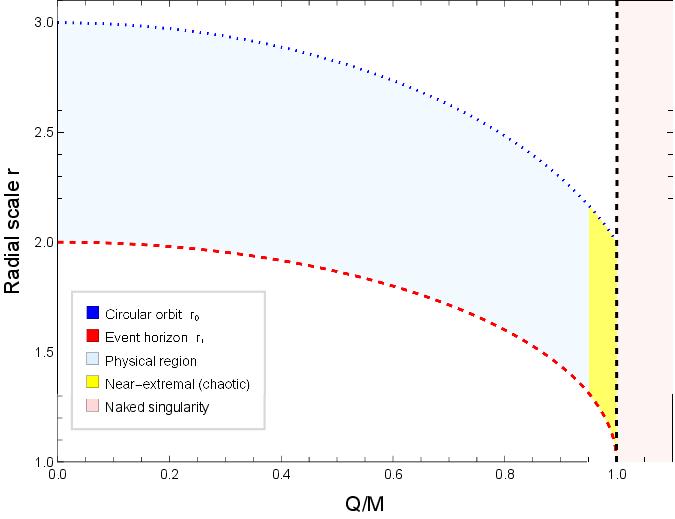}
\\ (d)
\end{minipage}

\caption{(a) Phase structure of  black holes in the $(Q/M,\, r)$ plane.
The red solid curve denotes the event horizon radius $r_{+}$, while the blue dotted curve represents the circular orbits around the horizon.
The light-blue shaded region corresponds to physically admissible configurations in which the dynamical interaction between the orbiting particle and the black hole occurs.
The yellow shaded region indicates the near-extremal regime, where the horizon approaches extremality, and the particle orbits become chaotic before encountering the naked singularity (red shaded region).
The black dotted vertical line marks the boundary between extremal black holes and naked singularities.
Panel (a) shows the global projection of the dynamics in the shift-symmetric Horndeski black hole, which is enlarged in panel (b) for clarity. Panels (c) and (d) show the same mechanism for the EGB and the Reissner-Nordstr\"om black holes, respectively.
}
\label{Ph}
\end{figure}

\begin{table}[t]
\centering
\caption{
Chaos bound and its apparent violation for the  Reissner-Nordstr\"om black hole. Here $\lambda_{\rm time}$ denotes the Lyapunov exponent for neutral timelike orbits ($q=0$), evaluated at $\epsilon=10^{-4}$, with the corresponding angular momentum $L_{0}$ (relevant only for timelike trajectories). The quantity $\lambda_{\rm null}$ denotes the Lyapunov exponent associated with
null circular orbits. Positive values of $\lambda^{2}-\kappa^{2}$ signal apparent violations of the MSS chaos bound, while negative values indicate bound preservation.
}
\label{tab:RN-chaos}

\renewcommand{\arraystretch}{1.12}
\setlength{\tabcolsep}{6pt}
\small

\begin{tabular*}{\columnwidth}{@{\extracolsep{\fill}}cccccc}
\hline\hline

$Q/M$ &
$r_{+}$ &
$r_{0}$ &
$L_{0}$ &
$\lambda_{\rm time}^{2}-\kappa^{2}$ &
$\lambda_{\rm null}^{2}-\kappa^{2}$ \\

\hline

0.6000 & 1.80000 & 2.73693 & 268.288 & $-0.02268$ & $-0.02267$ \\

0.8000 & 1.60000 & 2.48489 & 240.492 & $-0.01658$ & $-0.01657$ \\

0.8500 & 1.52678 & 2.39722 & 231.597 & $-0.01312$ & $-0.01303$ \\

0.9000 & 1.43589 & 2.29373 & 221.762 & $-0.00765$ & $-0.00760$ \\

0.9500 & 1.31225 & 2.16708 & 210.992 &
$\phantom{-}0.002414$ &
$\phantom{-}0.002412$ \\

1.0000 & 1.00000 & 2.00000 & 200.010 &
$\phantom{-}0.031250$ &
$\phantom{-}0.031250$ \\

\hline\hline
\end{tabular*}
\end{table}

\subsection{The Bardeen black hole}
To further confirm the robustness of this mechanism,  we study the Bardeen black hole, which provides an example of a regular (nonsingular) black hole spacetime, in contrast to the Reissner-Nordstr\"om solution, which possesses a curvature singularity at the origin. This enables us to determine whether the apparent violations of  the MSS bound are associated with curvature singularities or instead arise solely from the extremal geometric configuration. This regularity makes it a useful geometry to consider when investigating the behavior of dynamical quantities near the would-be singular region of charged black holes. We consider the static and spherically symmetric Bardeen black hole solution introduced in Ref.~\cite{bardeen1968non}, for which the metric function appearing in Eq.~\eqref{hb} is given by
\begin{equation}
h(r)=f(r) = 1 - \frac{2 M r^2}{\left(r^2 + Q^2\right)^{3/2}},
\end{equation}
and the angular part is the same as the Reissner-Nordstr\"om's.
 In the asymptotic limit $r\to\infty$, the metric reduces to the Schwarzschild form
\begin{equation*}
f(r) \approx 1 - \frac{2M}{r}\,,
\end{equation*}
while near the origin, the spacetime behaves as
\begin{equation*}
f(r) \approx 1 - \frac{2M}{Q^3} r^2\,,
\end{equation*}
which remains finite and indicates the absence of a curvature singularity at $r=0$. The parameter $Q$ is interpreted as a magnetic charge arising in nonlinear electrodynamics, and the spacetime admits two horizons for sufficiently small $Q$, which merge into a degenerate (extremal) horizon at $Q = \frac{4M}{3\sqrt{3}}$ \cite{fernando2012quasinormal}.
\begin{figure}[t]
\centering

\begin{minipage}{0.48\textwidth}
\centering
\includegraphics[width=\textwidth]{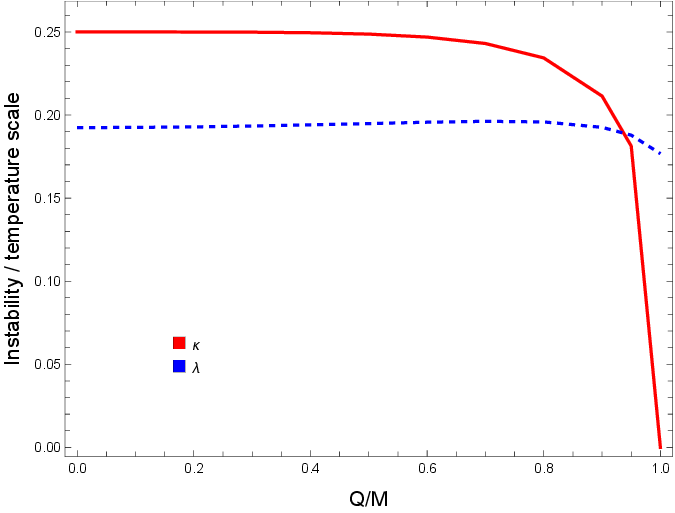}
\\ (a)
\end{minipage}
\hfill
\begin{minipage}{0.48\textwidth}
\centering
\includegraphics[width=\textwidth]{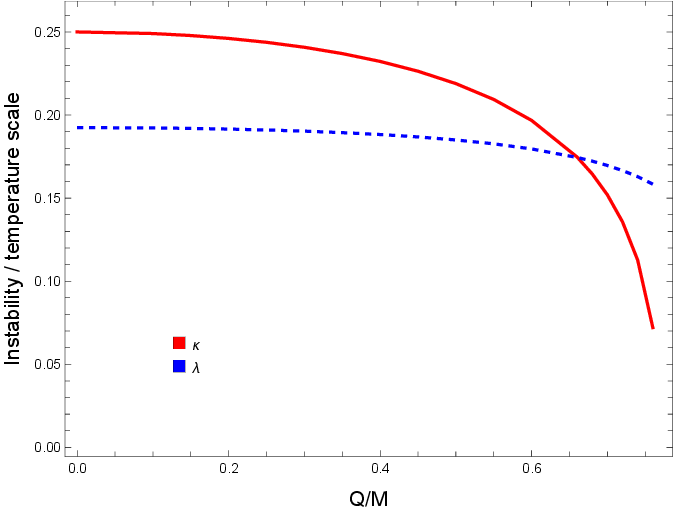}
\\ (b)
\end{minipage}
\hfill
\begin{minipage}{0.48\textwidth}
\centering
\includegraphics[width=\textwidth]{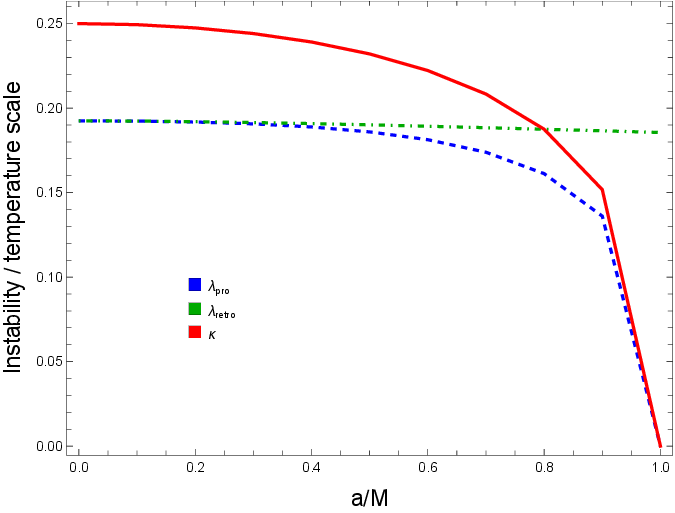}
\\ (c)
\end{minipage}

\caption{Instability and thermodynamic phase structure of different black holes. In all panels, the red curves represent the surface gravity, while the blue curves denote the corresponding Lyapunov exponent. Panels (a) and (b) show the Reissner-Nordstr\"om and Bardeen black holes, where the persistence of unstable circular orbits outside the degenerate horizon leads to apparent near-extremal violations of the MSS bound.  Panel (c) shows the Kerr black hole, where the blue (prograde) and green (retrograde) curves represent the two unstable circular orbits. The prograde orbit merges with the extremal horizon and saturates the bound, whereas the retrograde orbit remains outside the horizon, exhibiting the same apparent near-extremal violation observed in static black holes.}\label{tps}
\end{figure}

In this spacetime, our interest is primarily motivated by its distinct geometric structure, namely the absence of a curvature singularity. The location of the event horizon is determined by the largest root of the equation $h(r_+)=0$. Furthermore, by applying the photon sphere condition given in Eq.~\eqref{ps}, we find that the radius of the photon sphere satisfies
\begin{equation}
-3 M r_0^4 \sqrt{Q^2+r_0^2}+Q^6+3 Q^4 r_0^2+3 Q^2 r_0^4+r_0^6=0\,.
\end{equation}
The analysis of chaotic orbits is summarized in Table~\ref{bdt}, while the corresponding behavior is illustrated in Fig.~\ref{tps}(b). The results follow a pattern similar to that of the Reissner–Nordstr\"om case for large charge-to-mass ratios approaching extremality. The Bardeen black hole, even without a curvature singularity, exhibits the same qualitative trend: the surface gravity tends to zero while the Lyapunov exponent remains finite, thereby violating the chaos bound. For example, as the charge-to-mass ratio increases toward the extremal limit, the surface gravity tends to zero due to the emergence of a degenerate horizon, whereas the Lyapunov exponent remains finite because unstable circular orbits persist in this regime. This implies that the chaos-bound violations do not arise from curvature singularities but rather from the presence of a degenerate horizon, thereby revealing their fundamentally geometric nature.  

\subsection{Spinning black holes}
\label{secke}
Reference~\cite{kan2022bound} reported violations of the chaos bound for Kerr and Kerr-Newman black holes, particularly in the near-extremal regime. As we demonstrate below, these violations arise from the same geometric mechanism identified for static black holes, namely the decoupling of the thermal and instability scales near extremality. We therefore revisit the Kerr geometry to show that rotation does not alter the underlying origin of the anomaly.

The Kerr spacetime in Boyer-Lindquist \cite{boyer1967maximal} coordinates is described by
\begin{eqnarray}
ds^2&=&-\left(1-\frac{2Mr}{\Sigma}\right)dt^2
-\frac{4Mar\sin^2\theta}{\Sigma}\,dt\,d\phi
+\frac{\Sigma}{\Delta}dr^2+\Sigma\,d\theta^2
\nonumber\\
&&+\left(r^2+a^2+\frac{2Ma^2r\sin^2\theta}{\Sigma}\right)
\sin^2\theta\,d\phi^2,
\end{eqnarray}
where
\begin{equation}
\Sigma=r^2+a^2\cos^2\theta,\qquad
\Delta=r^2-2Mr+a^2,
\end{equation}
$M$ and $a$ denote the black hole mass and spin parameter, respectively~\cite{kerr1963gravitational}. The outer and inner horizons are located at the roots of $\Delta=0$,
\begin{equation}
r_\pm=M\pm\sqrt{M^2-a^2},
\end{equation}
with extremality reached at $a=M$, where $r_+=r_-=M$.

A distinctive feature of rotating black holes is the existence of two unstable equatorial photon orbits: the inner prograde orbit ($r_0^-$) and the outer retrograde orbit ($r_0^+$). Unlike static black holes with two horizons, where the inner unstable orbit lies beneath the event horizon and is therefore inaccessible,
\begin{equation}
r_-<r_0^-<r_+<r_0^+,
\end{equation}
both Kerr photon orbits remain outside the event horizon throughout the subextremal regime,
\begin{equation}
r_+<r_0^-<r_0^+.
\end{equation}

The equatorial photon orbits satisfy~\cite{hioki2009measurement}
\begin{equation}
r_0^2-3Mr_0\pm2a\sqrt{Mr_0}=0,
\label{kob}
\end{equation}
whose closed-form solutions are
\begin{align}
r_0^-&=
2M\left\{
1+\cos\!\left[
\frac{2}{3}\cos^{-1}\!\left(-\frac{a}{M}\right)
\right]
\right\},\\
r_0^+&=
2M\left\{
1+\cos\!\left[
\frac{2}{3}\cos^{-1}\!\left(\frac{a}{M}\right)
\right]
\right\}.
\end{align} 

\begin{table}[t]
\centering
\caption{Apparent violations of the MSS chaos bound for null circular orbits around the Bardeen black hole at different charge-to-mass ratios. The apparent violation occurs for $Q/M\gtrsim0.68$.}
\label{bdt}

\renewcommand{\arraystretch}{1.12}
\setlength{\tabcolsep}{6pt}
\small

\begin{tabular*}{\columnwidth}{@{\extracolsep{\fill}}cccccc}
\hline\hline

$Q/M$ &
$r_{+}$ &
$r_{0}$ &
$\lambda$ &
$\kappa$ &
$\lambda^{2}-\kappa^{2}$ \\

\hline

0.50 & 1.78597 & 2.76871 & 0.185009 & 0.218916 & -0.0136959 \\

0.55 & 1.73143 & 2.71255 & 0.182670 & 0.209373 & -0.0104685 \\

0.60 & 1.66546 & 2.64674 & 0.179608 & 0.196752 & -0.0064525 \\

0.66 & 1.56421 & 2.55155 & 0.174478 & 0.174727 & -0.00008681 \\

0.68 & 1.52238 & 2.51481 & 0.172249 & 0.164550 & 0.00259312 \\

0.70 & 1.47436 & 2.47495 & 0.169654 & 0.151979 & 0.00568488 \\

0.72 & 1.41738 & 2.43141 & 0.166593 & 0.135693 & 0.00934088 \\

0.74 & 1.34540 & 2.38343 & 0.162921 & 0.112687 & 0.0138448 \\

0.76 & 1.23756 & 2.32991 & 0.158413 & 0.0720929 & 0.0198972 \\

\hline\hline
\end{tabular*}
\end{table}

Since both photon orbits lie outside the event horizon, they each possess a distinct instability timescale characterized by their corresponding Lyapunov exponent. For equatorial null geodesics in the Kerr spacetime, the Lyapunov exponent is given by
\begin{equation}
\lambda_{\pm}
=
\sqrt{\frac{R''(r_{0}^{\pm})}
{2\,\dot{t}^{\,2}(r_{0}^{\pm})}}\,,
\label{eq:kerrlambda}
\end{equation}
where the radial potential is
\begin{equation}
R(r)
=
\left(r^{2}+a^{2}-ab_{\pm}\right)^{2}
-
\Delta(r)\left(b_{\pm}-a\right)^{2},
\label{eq:Rpot}
\end{equation}
and
\begin{equation}
\dot{t}
=
\frac{(r^{2}+a^{2})
\left(r^{2}+a^{2}-ab_{\pm}\right)}
{\Delta(r)}
-
a(a-b_{\pm}).
\label{eq:tdot}
\end{equation}
The corresponding impact parameter is
\begin{equation}
b_{\pm}
=
\frac{
aM
\pm
r\sqrt{a^{2}-2Mr+2r^{2}}
}
{M-r},
\label{eq:bphoton}
\end{equation}
where the upper (lower) sign corresponds to the retrograde (prograde) photon orbit. The surface gravity of the Kerr black hole is
\begin{equation}
\kappa
=
\frac{r_{+}-r_{-}}
{2\left(r_{+}^{2}+a^{2}\right)}.
\label{eq:kerrkappa}
\end{equation}
\begin{table}[t]
\centering
\caption{Horizon dynamics and apparent violations of the MSS chaos bound for the Kerr black hole at different spin parameters. At $a/M=0.90$, the Lyapunov exponent associated with the retrograde orbit begins to exceed the surface gravity and continues to do so toward extremality, resulting in an apparent violation of the bound. The inner prograde orbit does not exhibit this behavior.}
\label{kerrtab}

\renewcommand{\arraystretch}{1.12}
\setlength{\tabcolsep}{6pt}
\small

\begin{tabular*}{\columnwidth}{@{\extracolsep{\fill}}cccccccc}
\hline\hline

$a/M$ &
$r_{+}$ &
$r_{-}$ &
$r_{0}^{-}$ &
$r_{0}^{+}$ &
$\lambda(r_{0}^{-})$ &
$\lambda(r_{0}^{+})$ &
$\kappa$ \\

\hline

0.50 & 1.86603 & 0.133975 & 2.34730 & 3.53209 & 0.185918 & 0.190089 & 0.232051 \\

0.60 & 1.80000 & 0.200000 & 2.18891 & 3.62985 & 0.181303 & 0.189278 & 0.222222 \\

0.70 & 1.71414 & 0.285857 & 2.01333 & 3.72535 & 0.173888 & 0.188409 & 0.208309 \\

0.80 & 1.60000 & 0.400000 & 1.81109 & 3.81876 & 0.161230 & 0.187496 & 0.187500 \\

0.90 & 1.43589 & 0.564110 & 1.55785 & 3.91027 & 0.136080 & 0.186549 & 0.151784 \\

1.00 & 1.00000 & 1.00000 & 1.00000 & 4.00000 & 0.000000 & 0.185577 & 0.000000 \\

\hline\hline
\end{tabular*}
\end{table}
A detailed numerical analysis of the horizon structure and chaos-bound behavior for the Kerr black hole is presented in Table~\ref{kerrtab}. We find that the prograde photon orbit satisfies the chaos bound throughout the entire parameter space, including the extremal limit, because it merges with the degenerate event horizon as extremality is approached. By contrast, the retrograde orbit satisfies the bound only away from extremality. In the near-extremal regime, it remains separated from the event horizon even as the surface gravity vanishes, giving rise to an apparent violation of the chaos bound. This is precisely the behavior reported in Ref.~\cite{kan2022bound}.

Beyond their role in the present discussion, the two orbits $r_0^{\mp}$ set observationally accessible quantities through the eikonal correspondence between unstable photon orbits and black hole
quasinormal modes \cite{cardoso2009geodesic,hod2011fastest,hod2018lower,yang2012quasinormal}: at large angular harmonic number $l$, the ringdown frequency and damping rate of the corotating (counter-rotating) branch of the quasinormal spectrum are controlled by the orbital frequency and Lyapunov exponent of the corresponding photon orbit, $\omega_{lmn}\approx m\,\Omega(r_0^\mp)-i(n+\tfrac12)\lambda(r_0^\mp)$. Table~\ref{kerrtab} therefore doubles as a prediction for the near-extremal ringdown spectrum: because $\lambda(r_0^-)\to0$ while $\lambda(r_0^+)$ remains finite as $a\to M$, the corotating branch of the quasinormal spectrum becomes undamped in the extremal limit, while the counter-rotating branch retains a finite decay rate. This is precisely the bifurcation into zero-damped and damped mode families identified in Refs. \cite{yang2013quasinormal, zimmerman2016damped} in near-extremal Kerr ringdown, and our results identify its geometric
origin: the zero-damped modes are the quasinormal echo of the branch that saturates the MSS bound, and the damped modes are the echo of the branch that appears to violate it.

Remarkably, this is the same geometric mechanism identified for static black holes with inner and outer horizons, demonstrating that the origin of these apparent violations is universal and largely independent of the underlying gravitational theory. The same conclusion applies to both null and massive geodesics, indicating that the phenomenon is governed by the relative evolution of the unstable orbit and the horizon structure rather than by the nature of the geodesic itself. Motivated by this universal behavior, we examine in the following section the general conditions under which the chaos bound is expected to be satisfied or apparently violated.

\subsection{Geometric criterion for the MSS bound}
\label{sec:geometriccriterion}
The near-extremal behavior of the Lyapunov exponent can be understood directly from the relative behavior of the unstable circular orbit and the event horizon. Consider a static and spherically symmetric family of geometries of the form given in Eq.~\eqref{hb}, with $h,f,D\in C^2\big([r_+(\mu),r_+(\mu)+\delta)\big)$ for some $\delta>0$ and $D[r_+(\mu)]>0$. For every non-extremal member $\mu\neq\mu_*$, let $h$ and $f$ share a simple zero at $r_+(\mu)$. Let $r_0(\mu)$ denote an unstable circular orbit and \(\kappa(\mu) = \frac{1}{2} \left.\sqrt{\frac{h(r)}{f(r)}}\,f'(r)
\right|_{r=r_+(\mu)} \)
the corresponding surface gravity. We assume that the family varies
continuously with $\mu$ and approaches an extremal configuration as
$\mu\to\mu_*$, such that
$\kappa(\mu)\to0$. The two possible behaviors of the orbit in this limit can then be seen directly by analyzing the general expression \eqref{lyapgeneral2}.\footnote{Here, $C^2$ denotes twice continuous differentiability; that is, $f$, $h$, and $D$, together with their first and second derivatives, are continuous on the specified interval.}

\begin{itemize}
    \item [(i)] The first possible case: suppose that the unstable orbit approaches the horizon in the extremal limit, 
\begin{equation}
r_0(\mu)\to r_+(\mu_*).
\end{equation}
From Eq.~\eqref{lyapgeneral2}, $C^2$ regularity and $D(r_+)>0$ bound the bracketed term as $r_0\to r_+$, while $f(r_0)\to f(r_+)=0$. Hence
\begin{equation}
\lambda^2\to0,
\end{equation}
and therefore
\(
\lambda\to0.
\)
Thus, when the unstable orbit itself is driven into the near-horizon region, the same geometric degeneration that drives $\kappa\to0$ also suppresses the Lyapunov exponent.

The behavior is fundamentally different if the unstable orbit remains outside the horizon as the extremal limit is approached, yielding another distinct case.
\item[(ii)] Suppose
\(
r_0(\mu)\to r_0^*>r_+(\mu_*)
\)
for fixed $r_0^*$. Continuity then gives $f(r_0)\to f(r_0^*)>0$ and a bounded, generically nonzero bracket, so that 
\begin{equation}
\lambda(\mu)\to\lambda^*
=
\sqrt{
f(r_0^*)
\left[
\frac{h(r_0^*)}{D(r_0^*)}
-\frac12h''(r_0^*)
\right]
}>0,
\end{equation}
while
\(
\kappa(\mu)\to0.
\)
Consequently,
\begin{equation}
\frac{\lambda(\mu)}{\kappa(\mu)}\to\infty.
\end{equation}
\end{itemize} 
These two limits capture the essential geometric distinction underlying the near-extremal behavior. If the unstable orbit follows the horizon into the degenerate region, the vanishing of $f(r_0)$ forces the instability scale to vanish as well. If instead the orbit remains at a finite distance outside the extremal horizon, its local instability remains finite even though the
surface gravity collapses. The Lyapunov and thermal scales therefore need not vanish together, and their separation provides the geometric origin of the chaos-bound violation discussed in this work.

The first case establishes only that $\lambda\to0$. The sharper statement $\lambda/\kappa\to1^-$ requires controlling the rate at which $r_0(\mu)-r_+(\mu)$ vanishes relative to $\kappa(\mu)$, which is fixed only once an equilibrium (geodesic or force-balanced) condition on the orbit is imposed. We establish this rate, and hence $\lambda/\kappa\to1^-$, via the
near-extremal double-scaling construction of Sec.~\ref{secc4}.
\subsubsection{Extension to rotating black holes}
\label{sec:rotatinggeneralization}

The same geometric argument extends naturally to stationary and axisymmetric black holes. Consider, for example, a Kerr-type geometry in Boyer--Lindquist coordinates, with the horizons determined by
\(\Delta(r_\pm)=0,\)
and the equatorial unstable circular photon orbits condition given in Eq. \eqref{kob}. In the rotating case, the coordinate-time factor entering the orbital dynamics takes the form
\begin{equation}
\dot t=\frac{N(r_0,a,E,L)}{\Delta(r_0)},
\end{equation}
where $N$ is a bounded, generically nonzero function encoding the energy-angular momentum coupling induced by the rotation of the spacetime.

As extremality is approached, the surface gravity
\begin{equation}
\kappa(\mu)
=
\frac{\Delta'[r_+(\mu)]}
{2\big[r_+(\mu)^2+a(\mu)^2\big]}\,,
\end{equation}
vanishes. Since the Kerr black hole possesses two physically unstable orbits, the same two geometric possibilities identified in the static geometries arise here as well, branch by branch. For a branch whose unstable circular orbit approaches the extremal horizon,
\begin{equation}
r_0(\mu)\to r_+(\mu_*),
\end{equation}
we have $\Delta(r_0)\to0$ and hence $\dot t\to\infty$. The corresponding instability measured in coordinate time is therefore suppressed, giving
\(
\lambda\to0.
\)
By contrast, if a branch of unstable circular orbits remains separated from the horizon,
\(
r_0(\mu)\to r_0^*>r_+(\mu_*),
\)
then $\Delta(r_0)$ remains finite and so does the coordinate-time conversion factor. The orbital instability consequently approaches a finite value,
\(
\lambda\to\lambda^*>0,
\)
while
\(
\kappa\to0,
~\frac{\lambda}{\kappa}\to\infty.
\)

Thus, rotation does not alter the underlying geometric mechanism. It changes the detailed orbital dynamics through frame dragging and produces distinct prograde and retrograde branches, but the decisive feature remains the relative location of the unstable orbit and the extremal horizon. An orbit that follows the horizon inherits the suppression associated with the degenerate geometry, whereas an orbit that remains outside the near-horizon region retains a finite instability while the surface gravity vanishes.

The mechanism responsible for the apparent violation of the MSS bound is fundamentally geometric and is therefore largely insensitive to the underlying gravitational theory. In both EGB and shift-symmetric Horndeski black holes, the same qualitative behavior emerges: as the extremal limit is approached, the surface gravity $\kappa$, and hence the Hawking temperature, collapses owing to the degeneracy of the event horizon, whereas the unstable circular geodesics remain outside the near-horizon throat and continue to possess a finite Lyapunov exponent. Consequently, the thermodynamic scale, characterized by $\kappa$, becomes decoupled from the dynamical instability scale governing the orbital motion. This near-extremal decoupling provides a natural geometric explanation for why apparent violations of the chaos bound arise systematically either in the extremal limit or beyond sufficiently large charge-to-mass ratios, as reported in Refs.~\cite{wang2023spatial,Targema2025MSS,kan2022bound,gwak2022violation,c2,xie2023circular,dalui2026chaotic,li2026chaotic}. Rather than reflecting a breakdown of the bound itself, these observations indicate that the quantities being compared originate from distinct geometric regions of the spacetime once extremality is approached.

This behavior has also been investigated from a broader perspective in Ref.~\cite{d2}, where a wide class of black hole spacetimes and orbital configurations was examined. With the exception of the near-extremal regime, unstable geodesics were found to satisfy the bound rather generally, lending further support to the geometric picture developed here. The missing ingredient, however, has been a unified explanation for why the near-extremal regime behaves differently. Our analysis identifies this mechanism as the geometric decoupling of the thermodynamic and dynamical instability scales near extremality. Motivated by this observation, we summarize the possible limiting behaviors of particle orbits before discussing their broader implications for black hole dynamics and the interpretation of the chaos bound.
In general, four qualitatively distinct situations may affect the behavior of $\lambda$ and $\kappa$ as summarized in the following cases:
\begin{itemize} 
\item[\textbf{Case 1.}] 
\textit{Black holes with lifted horizon degeneracy.} If an intrinsic geometric feature prevents the horizon from becoming degenerate anywhere in the physical parameter space, the near-extremal
mechanism responsible for apparent chaos-bound violations is absent, and the chaos bound remains satisfied throughout the parameter space. A representative example of this case is found in the torsion-corrected  Reissner-Nordstr\"{o}m black hole in Poincar\'e gauge gravity (see Appendix \ref{app:PGT} for more details, and for details about the geometry of the torsion-corrected Reissner-Nordstr\"om-like black hole, refer to Refs.  \cite{blagojevic2022entropy, cembranos2017new, Blagojevic2019Entropy}). In this model, the spacetime metric retains the same line element as the standard Reissner-Nordstr\"om solution, except that the charge parameter originates from torsion rather than a Maxwell field. In the torsion-dominated regime, the effective charge may take negative values, preventing the horizon from merging into an extremal state and thereby avoiding degeneracy. In this geometry and other geometries with similar characteristics, the chaos bound should be satisfied everywhere in the parameter domain.
\item[\textbf{Case 2.}] 
\textit{Regular and singular black holes with degenerate extremal regimes.}
For regular black holes that allow extremal states with degenerate horizons, the reasoning is similar to that of singular extremal solutions, indicating that the chaos bound may be violated in the near-extremal limit, as we demonstrated for the Bardeen and Reissner-Nordstr\"om-type black holes in this work. In this case, despite the absence of curvature singularities in the spacetime, a degenerate horizon renders the surface gravity vanishing, while unstable circular orbits exist, yielding a finite Lyapunov exponent and thus, an apparent violation of the MSS bound. The Hayward black hole \citep{hayward2006formation} represents a related case, as it supports an extremal configuration and exhibits behavior similar to the Bardeen solution, including the departure from the MSS bound in the near-extremal regime. 
\item[\textbf{Case 3.}] 
\textit{Black holes with a single horizon and without extremal or naked singularity regimes.}
For black holes that admit neither extremal configurations nor naked singularities, the horizon structure remains non-degenerate across the entire parameter space. In such geometries, unstable circular orbits exist only within a finite radial interval outside the horizon and disappear once the radius approaches the innermost stable circular orbit (ISCO). For example, in the Schwarzschild spacetime, unstable circular orbits occur for $3m \le r_0 < 6m$, while at $r_0=6m$ the orbit becomes marginally stable, and beyond this radius the orbits are stable. Since the unstable orbits are restricted to this finite region and never approach a regime where the surface gravity tends to zero, the associated Lyapunov exponent always remains bounded by the surface gravity. Consequently, the chaos bound is respected for all admissible unstable circular orbits. The Schwarzschild black hole, therefore, provides a canonical example of a geometry in which the bound is automatically satisfied due to the disappearance of unstable orbits at the onset of orbital stability.
\item[\textbf{Case 4.}]
\textit{Spinning black holes.}
Rotating black holes exhibit two distinct unstable equatorial photon orbits. In the Kerr spacetime, the inner prograde orbit approaches and eventually merges with the degenerate event horizon in the extremal limit, causing both the Lyapunov exponent and the surface gravity to vanish simultaneously. The MSS bound is therefore saturated along this branch. By contrast, the outer retrograde orbit remains outside the developing near-horizon throat as extremality is approached, so that its Lyapunov exponent remains finite while the surface gravity vanishes, leading to an apparent violation of the bound. Thus, the Kerr spacetime provides a particularly clear illustration of the proposed geometric criterion: whether the chaos bound is satisfied or apparently violated is determined solely by whether the relevant unstable orbit enters the near-horizon region in the extremal limit.
\end{itemize}

These cases, together with our geometric criterion, provide an explanation for the violations of the MSS bound reported in Refs.~\cite{d2,kan2022bound}. Moreover, it is important to emphasize that the mere presence of a cosmological constant is not sufficient to suppress chaotic behavior. Indeed, Refs.~\cite{Targema2025MSS,d2} demonstrated that black hole solutions with a cosmological constant, encompassing both de Sitter and anti-de Sitter geometries, can still exhibit apparent violations of the chaos bound at sufficiently large charge-to-mass ratios. However, if additional geometric or physical conditions are imposed on the parameter space—particularly in the presence of extra matter fields—the allowed domain may be further restricted in such a way that chaotic regimes are geometrically excluded (see Ref. \cite{Targema2025MSS} for relevant examples of this case).  These violations consequently reflect a universal aspect of extremal or near-extremal geometries rather than originating from the particle-model assumptions.  

The limiting cases discussed above demonstrate that the apparent near-extremal violations of the MSS bound do not signal a failure of the bound for black hole geodesics. Rather, the holographic interpretation of the MSS conjecture associates the relevant chaotic dynamics with the near-horizon region. Consequently, a meaningful comparison between the geodesic and OTOC Lyapunov exponents requires the unstable trajectories to probe the same near-horizon geometry, implying a synchronization between the instability scale and the thermodynamic scale up to extremality and within the $\mathrm{AdS}_2$ throat.

In this interpretation, the apparent anomaly should disappear once the unstable trajectories are localized within the near-horizon region. This suggests that black hole spacetimes may not intrinsically violate the MSS bound and that the observed apparent violations instead arise from a geometric mismatch between the orbital instability scale and the near-horizon thermal scale. Since this mismatch occurs only in the approach to extremality, restoring the near-horizon correspondence in the extremal regime removes the apparent violations throughout the parameter space considered.
The remainder of this section tests this hypothesis by considering charged particles, whose motion can be modified by electromagnetic interactions such that the instability is probed directly within the near-horizon region, including the extremal regime where the apparent anomaly arises.
\subsection{Charged-particle Lyapunov exponent in the throat}
\label{secc4}
To further examine the near-horizon origin of the chaos-bound behavior, we consider charged timelike particles whose radial motion can be controlled by electromagnetic interactions. Unlike null circular geodesics, these trajectories can be localized within the near-extremal throat without crossing the horizon, allowing a direct investigation of the instability scale in the region relevant to the MSS bound. We therefore analyze the Lyapunov exponent of charged particles in the near-horizon geometry, with particular attention to the role of angular momentum in the throat. A related near-horizon analysis of the MSS bound was presented in Ref.~\cite{hashimoto2017universality}. The key distinction is that our analysis focuses specifically on the near-horizon geometry of near-extremal black holes, where the emergence of the throat and the associated separation of the instability and thermal scales play a central role.

Since the mechanism identified above is tied to the universal geometric structure of near-extremal horizons rather than to the detailed form of the underlying gravitational theory, we use the Reissner-Nordstr\"om black hole as a concrete realization in which the near-horizon analysis can be performed explicitly. The results obtained below are therefore intended to illustrate the behavior expected for a broader class of near-extremal black holes whose throat geometries reduce to an AdS$_2\times S^2$-type structure. In particular, such throats are described at low energies by an effectively two-dimensional dilaton  gravity, with Jackiw-Teitelboim (JT) gravity providing the canonical near-extremal description. The Reissner-Nordstr\"om geometry thus serves here as an explicit representative of a more general near-extremal throat mechanism. 

Following Refs.~\cite{porfyriadis2019near,maldacena1999anti,spradlin1999vacuum,moitra2019jackiw,mertens2023solvable}, we first introduce the throat coordinates by defining
\(
\rho=r-r_+,
\)
and parameterizing the separation between the two horizons as
\begin{equation}
\label{b}
r_+-r_-=2b,
\end{equation}
for convenience. In this case, the near-extremal limit is obtained as $b\rightarrow0$. To probe the near-horizon region, we simultaneously take $\rho\rightarrow0$, with the radial position controlled by the electromagnetic interaction. By expanding the metric function around the outer horizon, we find
\begin{equation}
    f(r)=\frac{(r-r_+)(r-r_-)}{r^2}
    =
    \frac{\rho(\rho+2b)}{r_+^2}
    \left[
    1-\frac{2\rho}{r_+}
    +\frac{3\rho^2}{r_+^2}
    +\cdots
    \right].
\end{equation}
Thus, at leading order,
\begin{equation}
    f(\rho)
    =
    \frac{\rho(\rho+2b)}{r_+^2}
    +\mathcal{O}\left(
    \frac{\rho^3+b\rho^2}{r_+^3}
    \right).
\end{equation}
In this limit, the surface gravity is
\begin{equation}
    \kappa=\frac{b}{r_+^2},
    \label{fkappa}
\end{equation}
while the radial function becomes $D(r)=r_+^2$ at leading order in the near-extremal limit $r_+\rightarrow M$ and $b\rightarrow0$ (see Eq.~\eqref{rind}). This corresponds to the standard frozen-radius $AdS_2\times S^2$ throat geometry; the subleading corrections to $D(r)=(r_++by)^2$ are retained later (see Appendix \ref{app:exact-radial}) when the full geometry is restored.
We now introduce the rescaled throat coordinate
\begin{equation}
\label{ro}
    y\equiv\frac{\rho}{b},
\end{equation}
and take the double-scaling limit with $y$ held fixed. The metric function and its first two derivatives are given by
\begin{eqnarray}
    f&=&\frac{b^2}{r_+^2}y(y+2),\label{f}\\
    f'&=&\frac{2b}{r_+^2}(y+1),\label{f1}\\
    f''&=&\frac{2}{r_+^2},\label{f11}
\end{eqnarray}
where primes denote derivatives with respect to $\rho$.

In these coordinates, the coefficients entering the angular-momentum constraint, given in Eqs.~\eqref{coeff:alpha}--\eqref{coeff:gamma}, become
\begin{eqnarray}
    \mathcal{A}_1&=&\frac{4b^2}{r_+^8}(y+1)^2,\\
    \mathcal{A}_2&=&\frac{4b^2}{r_+^6}
    \left[2(y+1)^2-q^2y(y+2)\right],\\
    \mathcal{A}_3&=&\frac{4b^2}{r_+^4}
    \left[(y+1)^2-q^2y(y+2)\right].
\end{eqnarray}
The corresponding solutions for the angular momentum are
\begin{eqnarray}
    L_0^2&=&-r_+^2,\label{unL0}\\
    L_0^2&=&r_+^2
    \left[
    \frac{q^2y(y+2)}{(y+1)^2}-1
    \right].\label{phL0}
\end{eqnarray}
The solution in Eq.~\eqref{unL0} is incompatible with the physical requirements that $L_0$ be real and satisfy $L_0^2\geq0$ (see also Ref.~\cite{Targema2025MSS}) and will therefore be discarded. The second solution is physical only for a restricted range of the charge-to-mass parameter $q$. Rather than determining this range numerically, it is more instructive to determine the charge required for a stationary equilibrium at a prescribed location in the throat. The equilibrium condition, obtained from Eq.~\eqref{pdot}, is
\begin{equation}
    f'\left(1+\frac{L^2}{D}\right)-f\frac{L^2D'}{D^2}+2q\Phi_t'\sqrt{f\left(1+\frac{L^2}{D}\right)}=0,
\end{equation}
so that the equilibrium charge satisfy
\begin{equation}
    q=-\frac{f'}{2\Phi_t'\sqrt{f}}\sqrt{1+\frac{L^2}{r_+^2}}\,,\label{cc}
\end{equation}
where we have used $D=r_+^2$. Accordingly, the equilibrium dynamics in the throat may be analyzed from two complementary perspectives: allowing for nonvanishing angular momentum (circular motion) or considering the purely radial limit. As we show below, both descriptions lead to the same physical conclusion.
\subsubsection{Case A: Circular motion in the throat}
We first retain a nonvanishing angular momentum and consider the most general equilibrium configuration. Under these conditions, the Lyapunov exponent in Eq.~\eqref{eqn: Glyapunov} reduces to
\begin{equation}
    \lambda^2=\frac{f'^2}{4}-\frac{ff''}{2}-\frac{q\Phi''_tf^2}{\sqrt{f\left(1+\frac{L^2}{r_+^2}\right)}}\,.
\end{equation}
At equilibrium orbit, we use the exact charge in Eq. \ref{cc} and obtain
\begin{equation}
     \lambda^2=\frac{f'^2}{4}-\frac{ff''}{2}+\frac{f'\Phi''_tf}{2\Phi_t'}\,.
     \label{nl}
\end{equation}
Remarkably, after imposing the equilibrium condition, the explicit dependence on the angular momentum disappears identically. Thus, the Lyapunov exponent is determined entirely by the throat geometry.
\subsubsection{Case B: Radial motion in the throat}
The purely radial configuration is recovered by setting $L=0$ in Eq.~\eqref{cc}. Accordingly, the equilibrium condition reduces to
\begin{equation}
\label{echarge}
    -q\Phi_t'
    -\frac{f'}{2\sqrt{f}}
    =0,
\end{equation}
and the equilibrium charge satisfying this condition is obtained as 
\begin{equation}
\label{trotcharge}
    q(y)
    =
    -\frac{f'(\rho_0)}
    {2\Phi_t'(\rho_0)\sqrt{f(\rho_0)}},
\end{equation}
where $\rho_0$ denotes the equilibrium position in the rescaled geometry. In the throat, the electric potential gradient takes the constant value
\(
    \Phi_t'=-\frac{1}{r_+},
\)
so that $\Phi_t''=0$. 

We can now simplify the Lyapunov exponent in Eq.~\eqref{eqn: Glyapunov}. In the equilibrium configuration, we find that the Lyapunov exponent for radial geodesics reduces exactly to Eq.~\eqref{nl}. This demonstrates that, within the throat geometry, radial and circular motions exhibit the same physical behavior, which can therefore be captured directly by analyzing Eq.~\eqref{nl}. Furthermore, since the throat electric field is constant, $\Phi_t''=0$, Eq.~\eqref{nl} simplifies to
\begin{equation}
    \lambda^2
    =
    \frac{f'^2}{4}
    -\frac{ff''}{2}.
\end{equation}
By considering Eqs.~\eqref{f}--\eqref{f11}, we obtain
\begin{equation}
    \lambda^2
    =
    \frac{b^2}{r_+^4}(y+1)^2
    -
    \frac{b^2}{r_+^4}y(y+2)
    =
    \frac{b^2}{r_+^4}.
    \label{yLyap}
\end{equation}
Finally, using Eq.~\eqref{fkappa}, we find
\(
    \lambda^2=\kappa^2.
\)
Thus, for a charged-particle trajectory localized within the near-extremal throat of the static black hole geometry, the geodesic Lyapunov exponent exactly saturates the MSS bound,
\(
    \lambda=\kappa.
\)
Importantly, this saturation persists in the extremal limit. As $b\rightarrow0$, both the Lyapunov exponent and the surface gravity vanish, $\lambda,\kappa\rightarrow0$, while their equality remains exact. The extremal throat therefore represents the smooth zero-temperature limit of the MSS-saturating relation.
Notably, the analysis is independent of the choice of time coordinate. Introducing the dimensionless time variable $\tau=\kappa t$, the throat metric can be written as
\begin{equation}
\label{eads}
    \frac{ds^2}{r_+^2}= -y(y+2)d\tau^2+\frac{dy^2}{y(y+2)}+d\Omega^2.
\end{equation}
By repeating the derivation with $\tau$ as the evolution parameter instead of the Schwarzschild-like time $t$ reproduces all of the preceding results exactly. Importantly, this result demonstrates that once the unstable trajectory is localized within the throat, the geodesic Lyapunov exponent exactly saturates the MSS bound, even arbitrarily close to extremality. It follows that the apparent violations observed in the near-extremal regime, both in the present work and in several previous studies, should be understood as a geometric anomaly arising from the inability of unstable circular geodesics to probe the near-horizon region where the holographic description is expected to apply. Although the broader question of whether the geodesic Lyapunov exponent and the OTOC Lyapunov exponent are universally identical remains open, the present analysis shows that, in the absence of additional external effects that displace the relevant trajectories from the throat (such as particle spin as shown in Ref. \cite{hashimoto2017universality}), both obey the same universal bound and exhibit remarkably consistent behavior.

\subsubsection{Exact saturation in the quadratic throat}
\label{sec:quadraticthroat}
The near-extremal double-scaling limit allows the behavior found above to be sharpened further. The essential observation is that the resulting throat geometry has an exactly quadratic blackening function. Consider the static
geometry
\begin{equation}
ds^2=-F(y)\,d\tau^2+\frac{dy^2}{F(y)}+D_0\,d\Omega^2,
\end{equation}
with fixed transverse factor $D_0$ and
\begin{equation}
F(y)=\alpha_1 y^2+\beta_1 y,
\end{equation}
where the horizon is located at $y=0$. The corresponding surface gravity is
\begin{equation}
\kappa=\frac{F'(0)}{2}=\frac{\beta_1}{2}.
\end{equation}

For a probe held at any fixed $y$, whether by geodesic motion or by force balance, the Lyapunov exponent takes the form
\begin{equation}
\lambda(y)^2
=
\frac{F'(y)^2}{4}
-\frac{F(y)F''(y)}{2}.
\end{equation}
Using
\begin{equation}
F'=2\alpha_1 y+\beta_1,
\qquad
F''=2\alpha_1,
\end{equation}
one finds
\begin{align}
\frac{F'^2}{4}-\frac{FF''}{2}
&=
\frac{(2\alpha_1 y+\beta_1)^2}{4}
-\alpha_1(\alpha_1 y^2+\beta_1 y)
\nonumber\\
&=
\alpha_1^2y^2+\alpha_1\beta_1 y+\frac{\beta_1^2}{4}
-\alpha_1^2y^2-\alpha_1\beta_1 y
\nonumber\\
&=
\frac{\beta_1^2}{4}
=
\kappa^2.
\end{align}
Therefore,
\begin{equation}
\lambda(y)^2=\kappa^2,
\end{equation}
identically and independently of $y$. Thus, once the dynamics is governed by the quadratic throat geometry, the Lyapunov exponent saturates the MSS bound everywhere in the throat, rather than only asymptotically at the horizon.

This result applies directly to the near-extremal throat obtained in Sec.~\ref{secc4}. In units with $r_+=1$, the throat metric \eqref{eads} has
\begin{equation}
F(y)=y(y+2),
\end{equation}
with $\tau=\kappa t$, corresponding to
\begin{equation}
\alpha_1=1,\qquad \beta_1=2.
\end{equation}
The appearance of this quadratic form is not accidental. In the double-scaling limit, the generic, non-quadratic blackening function $f(r)$ reduces, at leading order in $b$, to an exactly quadratic function of the throat coordinate $y$. This is the defining feature of the Rindler--AdS$_2$ slicing relevant to the near-extremal geometry.

The above identity therefore allows us to write that \(\lambda(y)=\kappa,\)
for every $y=\mathcal{O}(1)$ within the throat, rather than only in the strict near-horizon limit. The equilibrium charge in Eq.~\eqref{cc} provides the mechanism that places a physical probe at a chosen finite value of $y$. Accordingly, the explicit cancellation obtained in Eq.~\eqref{yLyap} is not an isolated feature of the particular equilibrium point selected by force balance; it is the $y$-independent value $\lambda^2=\kappa^2$ dictated by the quadratic throat geometry.

This establishes the sharper result
\begin{equation}
\frac{\lambda}{\kappa}\to1^-,
\end{equation}
for probes that remain at finite $y$ in the double-scaled throat. In this sense, the exact saturation of the chaos bound is a direct consequence of the universal quadratic structure of the near-extremal throat.

\subsubsection{Restoring subleading corrections to the throat result}
\label{sssec:exact-throat}
The preceding analysis shows that, within the leading-order quadratic throat geometry, the charged-particle Lyapunov exponent exactly saturates the MSS bound, $\lambda/\kappa=1$. This result, however, follows after truncating the full blackening function at leading order in the near-extremal expansion. We now retain the full Reissner--Nordstr\"om geometry and use the exact expression derived in Appendix~\ref{app:exact-radial} to determine how the instability-to-thermal ratio behaves beyond the quadratic approximation.

In this analysis, we focus on the radial equilibrium branch, $L=0$. This choice is motivated by the structure of the full problem rather than being a mere simplifying assumption. In the complete spacetime geometry, the equilibrium angular momentum $L_0$ is determined by the corresponding equilibrium condition, and our analysis shows that the angular momentum
sector does not introduce an independent mechanism for the apparent chaos bound violations (see Table \ref{tab:RN-chaos} for instance); the violations arise only in the near-extremal
regime. Having identified this regime as the relevant one, we turn to its near-horizon throat, where, as demonstrated in Sec.~\ref{secc4}, the circular and radial equilibrium configurations considered in Cases A and B yield the same Lyapunov exponent after the equilibrium condition is imposed. The radial branch therefore provides a direct description of the relevant instability scale in the throat without altering the physical conclusion.

As shown in Appendix~\ref{app:exact-radial}, the exact ratio of the instability and thermal scales is
\begin{equation}
\frac{\lambda}{\kappa}
=
\frac{1}
{\left(1+\dfrac{b}{r_+}y\right)^2}.
\label{exact_ration}
\end{equation}
Unlike the quadratic-throat result, Eq.~\eqref{exact_ration} retains the dependence on the terms beyond the leading-order near-horizon expansion and therefore allows us to examine the
instability-to-thermal ratio throughout the physical portion of the
throat.

For convenience, we write $x\equiv b/r_+$; since $r_-\ge0$ requires $b=(r_+-r_-)/2\le r_+/2$, the extremal-to-Schwarzschild family spans $x\in(0,\tfrac12]$ in general. The charged-particle construction used here, however, does not itself reach the full range: the equilibrium charge required to hold the probe at throat depth $y$ is given in Eq. \eqref{trotcharge}, obtained from the equilibrium condition in Eq. \eqref{echarge} at leading order in $x$. Since $\Phi_t'\propto Q$, the electric field varnishes as $Q\to0$: with no electric field, no Lorentz force can hold the probe in place, and the Schwarzschild endpoint $x=\tfrac12$ is excluded. The physical range for this construction is therefore the open interval $x\in(0,\tfrac12)$.  For any equilibrium charged-particle trajectory, $y\ge0$, yielding
\begin{equation}
    1+xy\ge1,
\end{equation}
with equality if and only if $y=0$. Equation~\eqref{exact_ratio} therefore gives
\begin{equation}
    \lambda\le\kappa,
    \qquad
    \text{for every } x\in(0,\tfrac12)\text{ and every }y\ge0.
    \label{eq:exactbound}
\end{equation}
Moreover, $\lambda/\kappa=(1+xy)^{-2}$ is strictly monotonically decreasing in $y$ for any fixed $x>0$, so there is no subleading
regime, and no accessible temperature within this construction in which the radially confined charged probe exceeds the surface gravity. The near-extremal, leading-order result $\lambda\to\kappa$ established in Sec.~\ref{secc4} is recovered as the $y\to0$ (equivalently $x\to0$) limit of the exact bound~\eqref{eq:exactbound}, not as an independent leading-order coincidence: saturation is the boundary case of a global inequality that holds identically across the accessible non-extremal-to-extremal family.

We note that Eq.~\eqref{trotcharge} gives $q>1$ for every $y>0$, approaching $1$ only as $y\to\infty$ and diverging as $y\to0$. Thus, in geometrized units, the probe is necessarily superextremal, with its charge-to-mass ratio exceeding unity. This is not unusual for a test particle; for example, the electron has $q/m\simeq\mathcal{O}(10^{21})$ in geometrized units. We nevertheless record this explicitly and, throughout, neglect the probe's self-force and backreaction on the background geometry.
\subsubsection{Why the photon sphere does not probe the quadratic throat}
\label{sec:photonspherethroat}
The above saturation result does not, however, apply to the photon sphere discussed in Case 2. The distinction is purely geometric. A neutral orbit held at a fixed physical radius $r_0$, with
\(
r_0-r_+=\mathcal{O}(1),
\)
independent of the near-extremal parameter $b$, corresponds in the throat coordinates to
\begin{equation}
y=\frac{r_0-r_+}{b}\to\infty,
\qquad (b\to0).
\end{equation}
Thus, although the black hole develops a quadratic near-horizon throat, the photon sphere is pushed to $y\to\infty$ in the throat coordinates and does not remain within the finite-$y$ region described by the double-scaled geometry.

Consequently, the photon sphere is governed by the full, non-quadratic blackening function $f(r)$ rather than by the truncated quadratic form $F(y)$ above. This is precisely the situation encountered in Case 2. The photon sphere is neutral and therefore has no mechanism to hold it at a finite value of $y$ as the throat develops. Its Lyapunov exponent must therefore be computed from the full geometry and, generically, approaches the finite, $\kappa$-independent value found in Case 2 rather than the
saturated value $\lambda=\kappa$. Accordingly, the resolution of the apparent anomaly lies in how the near-extremal limit is approached. If one treats extremal black holes as the limiting case of the general family and evaluates the geodesic instability outside the throat, the thermodynamic and dynamical scales decouple, producing an apparent violation. However, if the near-extremal geometry is instead treated as a distinct entity with its own near-horizon structure, and the orbit is properly localized within the AdS$_2$ throat, the instability scale tracks the surface gravity, and the bound is saturated. The anomaly is therefore not a failure of the MSS bound but an artifact of evaluating the classical probe outside the region where the holographic correspondence applies.

\subsection{Near-extremal OTOC from the Jackiw--Teitelboim throat}
\label{secjt}
The preceding sections establish that a suitably localized geodesic Lyapunov exponent equals $\kappa$ in the throat. We now show, rather than infer by analogy, that $\kappa$ is also the growth rate of the genuine OTOC (in the quantum-chaos sense) in this geometry—the quantity the MSS bound actually constrains—and that the photon-sphere exponent of Case 2 does not compute it.

\subsubsection{Reduction to Jackiw--Teitelboim gravity}
\label{sssec:JT}
In the double-scaling limit introduced in Sec.~\ref{secc4}, the near-horizon geometry reduces, at leading order in $b/r_+$, to $AdS_2\times S^2$ with metric \eqref{eads} and constant transverse
radius $r_+$. This is the Robinson-Bertotti geometry. At this order, the two-dimensional sector has the enhanced $SO(2,1)$ symmetry of $AdS_2$, while the leading departure from extremality is encoded in the variation of the transverse sphere. 
Indeed, since
\(
r=r_++\rho=r_++by,
\)
the transverse metric function is
\begin{equation}
\Psi(r)\equiv D(r)
=r_+^2+2r_+\phi(y)+\mathcal{O}(b^2),
\qquad
\phi(y)=by.
\end{equation}
Thus, the transverse area separates into its extremal value $\Psi_0=D(r_+)=r_+^2=M^2$, as used in Sec.~\ref{secc4}, and a leading fluctuation $\phi(y)$. 

Upon spherical reduction of the 4D-Einstein-Maxwell theory and expansion about the extremal throat,  the leading dynamical sector is described by JT gravity \cite{jackiw1985lower,teitelboim1983gravitation},
\begin{equation}
I_{\rm JT}
=
-\frac{1}{16\pi G_4}
\int d^2x\,\sqrt{g}\,
\psi\left(R_2+\frac{2}{r_+^2}\right)
-\frac{1}{8\pi G_4}
\oint dx\,\sqrt{\gamma}\,\psi_b K ,
\end{equation}
up to the standard boundary counterterm, where
\begin{equation}
\psi(y)\equiv\Psi(y)-\Psi_0=2r_+\phi(y),
\end{equation}
is the dynamical dilaton fluctuation, related to the radial offset $\phi(y)$ introduced above by the factor $2r_+$ fixed by expanding $\Psi(y)=D(r)$. This is the standard JT description of the near-extremal Reissner-Nordstr\"om throat \cite{moitra2019jackiw,porfyriadis2019near,maldacena2016remarks,
kitaev2018soft}, and it inherits the near-$AdS_2$/near-CFT$_1$
holographic description of
Refs.~\cite{almheiri2015models,maldacena2016conformal},  independent of the asymptotic structure of the full four-dimensional geometry: the OTOC computed below is therefore a genuine boundary correlator of the throat's own dual quantum mechanics, not a geometric analogue of one.

The normalization of the throat time relative to the asymptotic Killing time is important for identifying the physical temperature. Near the horizon at $y=0$, introduce $\eta=\sqrt{2y}$, so that
$y=\eta^2/2$, $dy=\eta\,d\eta$, and $y(y+2)=\eta^2(1+\eta^2/4)$. Hence, as $\eta\to0$, the throat metric \eqref{eads} becomes
\begin{equation}
\frac{ds^2}{r_+^2}
\longrightarrow
-\eta^2d\tau^2+d\eta^2+d\Omega^2.
\label{rindll}
\end{equation}
After Wick rotation, $\tau=-i\tau_E$, regularity at $\eta=0$ requires $\tau_E\sim\tau_E+2\pi$. Using the near-horizon scaling $\tau=\kappa t$ and identifying the boundary time with the original Killing time, $u=t$, this periodicity gives $\beta=2\pi/\kappa$ in the physical time
$u$, and therefore
\begin{equation}
T_H=\frac{1}{\beta}=\frac{\kappa}{2\pi}.
\label{eq:JT-temperature}
\end{equation}Thus, the unit surface gravity of the rescaled throat geometry corresponds, after restoring the normalization of the boundary time, to the physical Hawking temperature of the near-extremal black hole.
\subsubsection{The Schwarzian mode and its OTOC}
Because JT gravity has no local bulk gravitational degrees of freedom, its low-energy dynamics is captured by a boundary reparametrization mode. The map $\tau(u)$ between the physical boundary time $u$ and the fixed AdS$_2$ time $\tau$ of \eqref{eads} is governed by the Schwarzian action
\begin{equation}
    I_{\rm Sch}=-C_0\int du\,\mathrm{Sch}(\tau,u),
    \qquad
    \mathrm{Sch}(\tau,u)=\frac{\tau'''}{\tau'}-\frac{3}{2}\Big(\frac{\tau''}{\tau'}\Big)^2,
\end{equation}
with coupling $C_0$ fixed by the near-extremal specific heat of the Reissner-Nordstr\"om black hole and related to the black hole entropy as $S(T)=S_0+4\pi^2C_0\,T$ ~\cite{maldacena2016conformal}. The OTOC of boundary operators dual to bulk fields propagating in this throat is the standard Schwarzian OTOC computed in Refs.~\cite{maldacena2016conformal, jensen2016chaos}: at times $1\ll t\ll t_*$ (before the scrambling time $t_*\sim\beta\ln C_0$),
\begin{equation}
    \langle[W(t),V(0)]^2\rangle_\beta
    \propto\,e^{2\pi T_H t},
\end{equation}
so that
\begin{equation}
    \lambda_{\rm OTOC}=2\pi T_H=\kappa,
    \label{eq:lambdaOTOC}
\end{equation}
identically saturates the MSS bound. 

We now have three independently computed exponents in the same geometry:
\begin{align}
    \lambda_{\rm OTOC}\ \text{(Schwarzian/JT, Eq.~\eqref{eq:lambdaOTOC})}
        &=2\pi T_H=\kappa, \\
    \lambda_{\rm geo}\ \text{(throat-localized charged particle, Eq.~\eqref{eq:exactbound})}
        &\leq\kappa, \\
    \lambda_{\rm geo}\ \text{(static, uncharged photon sphere, Case 2)}
        &=\lambda^*>\kappa\, \ (\text{finite, }\kappa\text{-independent as }b\to0).
\end{align}
The first two agree exactly, by construction: both are properties of the
same quadratic throat geometry shown in Sec. \ref{sec:quadraticthroat}. The third does not track $\kappa$ because the photon-sphere orbit never enters the region -- the throat -- where either computation is defined: it sits at $y\to\infty$ in throat coordinates, outside the domain on which the JT reduction  apply (see Sec. \ref{sec:photonspherethroat}). 
\section{Conclusions}
\label{sec5}
This work presents a systematic investigation of the MSS chaos bound in black hole spacetimes using Lyapunov exponents across a broad class of gravitational theories. We examined shift-symmetric Einstein-Horndeski and Einstein-Gauss-Bonnet black holes together with the general relativistic Reissner-Nordstr\"om, Bardeen, and Kerr solutions. Across all spacetimes considered, we found that apparent violations of the MSS bound occur systematically in the near-extremal regime, indicating that the phenomenon is not specific to any particular gravitational theory or modified interaction.

The principal result of this work is the identification of the geometric origin of these apparent violations. We showed that the relevant criterion is whether the unstable trajectory probes the near-horizon throat that governs the holographic description. Whenever the trajectory remains localized within the throat, the instability scale tracks the surface gravity, yielding $\lambda=\kappa$, including in the extremal limit. When the unstable orbit remains outside the throat, however, the thermodynamic scale collapses while the orbital instability remains finite, producing the apparent near-extremal violations.

A direct verification of this interpretation was obtained by constructing charged-particle trajectories confined to the near-extremal throat. The electromagnetic interaction provides a controlled mechanism for localizing the unstable orbit arbitrarily close to the horizon, where the geodesic Lyapunov exponent saturates the MSS bound. This explicitly realizes, for the Reissner--Nordstr\"om geometry, the near-horizon universality discussed in Refs. \cite{hashimoto2017universality, d2,kan2022bound}, while extending the analysis to the double-scaled near-extremal throat and demonstrating that the same geometric criterion governs the broader class of black hole solutions considered here.

A second, independent verification follows from computing the OTOC directly in the near-extremal throat. Reducing the Einstein-Maxwell throat on the two-sphere yields JT gravity, whose boundary (Schwarzian) mode saturates the MSS bound identically, independent of the throat's overall normalization. This provides a genuinely quantum-mechanical—not merely geodesic—realization of
the saturating branch, and it agrees exactly with the charged particle in the throat dynamics, while the photon-sphere orbit responsible for the apparent violation sits outside the region where either calculation
applies.

A central lesson emerging from this work is that applying the MSS chaos bound to black holes requires more than simply computing a Lyapunov exponent. Within its holographic interpretation, the instability probe must remain tied to the same near-horizon geometry that determines the black hole temperature. For non-extremal black holes, the instability and thermodynamic scales remain synchronized throughout the parameter space we examined. However, as extremality is approached, this synchronization can break down. The development of the AdS$_2$ throat introduces a new geometric scale, allowing unstable geodesics to remain outside the throat while the thermal scale collapses with the surface gravity. It is precisely this geometric decoupling that gives rise to the reported apparent violations. Our analysis also shows that, although the near-extremal regime arises as the limiting case of generic black hole solutions, the limiting procedure alone is insufficient to capture the relevant instability dynamics. The reason is that the near-extremal AdS$_2$ throat constitutes a distinct geometric region whose dynamics in the context of MSS conjecture must be analyzed independently. When the instability is analyzed within this throat, the Lyapunov exponent again tracks the thermal scale, and the MSS bound is saturated. In this picture, non-extremal black holes satisfy the bound, while near-extremal and extremal black holes saturate it once the appropriate near-horizon dynamics are taken into account.

These results provide a unified interpretation of the reported violations of the MSS bound. Rather than signaling a breakdown of the bound itself, they arise because the instability probe becomes geometrically decoupled from the near-horizon thermal region that underlies the holographic description. Within this framework, the geodesic Lyapunov exponent provides a faithful classical probe of the MSS bound whenever the relevant unstable trajectory remains localized within the near-horizon throat. More generally, our results suggest that the relationship between classical instability and quantum chaos is governed not only by the existence of an unstable orbit, but also by whether that orbit probes the same near-horizon geometry that determines the thermodynamic sector of the black hole.

Finally, the geometric picture developed here naturally motivates extensions to higher-dimensional black holes, string-inspired geometries with additional degrees of freedom, and other gravitational theories in which the interplay between near-horizon geometry and orbital dynamics may differ from that of the black hole spacetimes considered here. 
\section*{Acknowledgements}
Terkaa Victor Targema and Usman Zafar acknowledge the Japanese government (MEXT) scholarship. The work of Kazuharu Bamba was supported in part by the JSPS KAKENHI Grants No.~24KF0100, No.~25KF0176, Competitive Research Funds for Fukushima University Faculty (No.25RK011), and a grant-in-aid of academic research of the Yamaguchi Scholarship Foundation. This work was also supported from the Research Promotion Project of Fukushima University Fund.

\section*{Data availability} 
The numerical code used to generate the results presented in this work is publicly available at \url{https://doi.org/10.5281/zenodo.22059578}.
\appendix

\section{Exact radial result in the Reissner-Nordstr\"om geometry}
\label{app:exact-radial}
By using the  parameterization of the separation between the two horizons introduced in Eq. \eqref{b}, it follows that
\begin{equation}
b=\sqrt{M^2-Q^2},
\qquad
M=\frac{r_++r_-}{2},
\qquad
Q^2=r_+r_-.
\end{equation}
The exact surface gravity is
\begin{equation}
\kappa
=
\frac{f'(r_+)}{2}
=
\frac{r_+-r_-}{2r_+^2}
=
\frac{b}{r_+^2}.
\label{exact_kappa}
\end{equation}

For radial motion, $L=0$, the equilibrium condition gives
\begin{equation}
q
=
-\frac{f'}{2\Phi_t'\sqrt{f}}.
\end{equation}
Substituting this condition into the general expression for the
Lyapunov exponent yields
\begin{equation}
\lambda^2
=
\frac{f'^2}{4}
-\frac{ff''}{2}
+\frac{f'\Phi_t''f}{2\Phi_t'}.
\label{exact_radial_lyap}
\end{equation}
For the Reissner-Nordstr\"om electromagnetic potential, $\Phi_t'=-Q/r^2$. Hence,
\begin{equation}
\frac{\Phi_t''}{\Phi_t'}=-\frac{2}{r},
\end{equation}
and Eq.~\eqref{exact_radial_lyap} becomes
\begin{equation}
\lambda^2
=
\frac{f'^2}{4}
-\frac{ff''}{2}
-\frac{ff'}{r}.
\label{exact_radial_lyap_RN}
\end{equation}

The derivatives of the exact metric function are
\begin{equation}
f'(r)
=
\frac{2M}{r^2}
-\frac{2Q^2}{r^3},
\end{equation}
and
\begin{equation}
f''(r)
=
-\frac{4M}{r^3}
+\frac{6Q^2}{r^4}.
\end{equation}
The three contributions in Eq.~\eqref{exact_radial_lyap_RN} are given by
\begin{align}
\frac{f'^2}{4}
&=
\frac{M^2}{r^4}
-\frac{2MQ^2}{r^5}
+\frac{Q^4}{r^6},
\\[4pt]
-\frac{ff''}{2}
&=
\frac{2M}{r^3}
-\frac{4M^2+3Q^2}{r^4}
+\frac{8MQ^2}{r^5}
-\frac{3Q^4}{r^6},
\\[4pt]
-\frac{ff'}{r}
&=
-\frac{2M}{r^3}
+\frac{4M^2+2Q^2}{r^4}
-\frac{6MQ^2}{r^5}
+\frac{2Q^4}{r^6}.
\end{align}
Adding these terms gives the exact simplification
\begin{equation}
\lambda^2
=
\frac{M^2-Q^2}{r^4}=\frac{b^2}{r^4}.
\label{exact_lambda_MQ}
\end{equation}
\begin{table*}[t]
\centering
\caption{Chaos-bound behavior in the torsion sector of Poincar\'e gauge gravity. The physical torsion sector, $q<0$, remains non-extremal and exhibits no apparent violation of the chaos bound. For comparison, the $q>0$ configurations correspond to the non-extremal Reissner-Nordstr\"om sector and likewise remain consistent with the bound.}
\label{tab:torsion_chaos}

\renewcommand{\arraystretch}{1.12}
\setlength{\tabcolsep}{6pt}
\small

\begin{tabular}{
c c c
@{\hspace{0.40cm}}
c c c
}
\hline\hline

$q$ &
$r_{+}$ &
$r_{0}$ &
$\lambda$ &
$\kappa$ &
$\lambda/\kappa$
\\

\hline

$0.001$ &
1.99950 &
2.99933 &
0.192461 &
0.250000 &
0.769843 \\

$0.200$ &
1.89443 &
2.86015 &
0.194442 &
0.249224 &
0.780191 \\

$-1.000$ &
2.41421 &
3.56155 &
0.181163 &
0.242641 &
0.746632 \\

$-10.000$ &
4.31662 &
6.21699 &
0.128346 &
0.177995 &
0.721065 \\

$-100.000$ &
11.0499 &
15.7215 &
0.0585412 &
0.0823087 &
0.711239 \\

\hline\hline
\end{tabular}

\end{table*}

Comparing Eq.~\eqref{exact_lambda_MQ} with the exact surface gravity
in Eq.~\eqref{exact_kappa},
we find
\begin{equation}
\frac{\lambda^2}{\kappa^2}
=
\frac{r_+^4}{r^4}
=
\left(\frac{r_+}{r}\right)^4,
\end{equation}
or, equivalently, \(\frac{\lambda}{\kappa}
=
\left(\frac{r_+}{r}\right)^2.
\label{exact_ratio_r}
\)
Thus the exact radial Lyapunov exponent can be written in the particularly
simple form
\begin{equation}
\lambda
=
\kappa\,\frac{r_+^2}{r^2}.
\label{exact_lambda_kappa}
\end{equation}

We can now introduce the throat coordinates as used in Eqs. \eqref{b} and \eqref{ro}, we have 
\begin{equation}
r=r_++\rho=r_++by
=r_+\left(1+\frac{b}{r_+}y\right).
\end{equation}
By substituting this relation into Eq.~\eqref{exact_ratio_r}, we obtain
\begin{equation}
\frac{\lambda}{\kappa}
=
\frac{1}
{\left(1+\dfrac{b}{r_+}y\right)^2}\,.
\label{exact_ratio}
\end{equation}
Equation~\eqref{exact_ratio} is an exact result for the radial
equilibrium configuration in the full Reissner--Nordstr\"om geometry;
no throat expansion has been used in obtaining it.

\section{Chaos bound in the torsion sector of Poincar\'e gauge gravity}
\label{app:PGT}
We briefly examine the chaos-bound behavior of the black hole solution arising in the torsion sector of Poincar\'e gauge gravity. The metric function is
\begin{equation}
f(r)=h(r)=1-\frac{2m}{r}+\frac{q}{r^2},
\end{equation}
where the torsion sector corresponds to $q<0$. In this sector, the horizon cannot become degenerate within the physical parameter domain, and hence the near-extremal geometric mechanism responsible for the apparent chaos-bound violations discussed in the main text is absent. As a result, the chaos bound remains satisfied throughout the physical
torsion sector, as shown in Table \ref{tab:torsion_chaos}.
\bibliographystyle{apsrev4-2}
\bibliography{apssamp}
\end{document}